\documentclass[12pt]{article}
\usepackage[letterpaper]{geometry}
\usepackage{amsmath,amssymb,amsthm,bm,graphicx,float,array,multirow,multicol,rotfloat,caption,subcaption,hyperref,cleveref,enumerate,mathdots,adjustbox,booktabs,parskip,mathtools,tikz,tikz-cd,pdflscape,csquotes,lscape,rotating,empheq,mathrsfs,longtable}
\usepackage[para]{threeparttable}
\usepackage[all]{xy}
\usepackage[normalem]{ulem}
\usepackage[aligntableaux=center]{ytableau}
\usepackage[numbers,sort&compress]{natbib}

\numberwithin{equation}{section}
\numberwithin{figure}{section}
\allowdisplaybreaks
\makeatletter
\usetikzlibrary{arrows,shapes.misc,positioning,decorations.pathmorphing,decorations.markings,decorations.pathreplacing,matrix,patterns,backgrounds}

\definecolor{gold}{RGB}{255,215,0}
\definecolor{purple}{RGB}{160,32,240}
\tikzset{
scale cd/.style={every label/.append style={scale=#1}, cells={nodes={scale=#1}}},
gauge/.style={rounded rectangle, draw=black!100, thick, minimum size=5mm}, 
gaugeD/.style={rounded rectangle, draw=black!100,double,thick,minimum size=5mm},  
empty/.style={rounded rectangle, draw=white!100, thick, minimum size=5mm}, 
flavor/.style={rectangle, draw=black!100, thick, minimum size=5mm},
flavorD/.style={rectangle, draw=black!100, double,thick, minimum size=5mm},
node/.style={circle, thick, draw=black!100,fill=white!100,  minimum size=2mm, inner sep=0pt},
sqnode/.style={rectangle
, thick, draw=black!100,fill=white!100,  minimum size=2mm, inner sep=0pt
},
sonode/.style={circle, thick, draw=black!100,fill=red!100,  minimum size=3mm, inner sep=0pt},
spnode/.style={circle, thick, draw=black!100,fill=blue!100,  minimum size=3mm, inner sep=0pt},
fnode/.style={rectangle, thick, draw=black!100,fill=white!100,  minimum size=3mm, inner sep=0pt},
tnode/.style={rounded rectangle, outer sep=0pt, thick, minimum size=5mm},
brace/.style={decoration={brace, mirror},decorate}
}

\newcommand\notsoscript{\@setfontsize\notsoscript{9}{7}}

\theoremstyle{plain}
\newtheorem*{thm*}{Theorem}

\theoremstyle{definition}

\newtheorem*{defn*}{Definition}

\makeatother

\def\scheme/{our affine scheme} 

\newcommand{\Crit}{\operatorname{Crit}}

\graphicspath{{figures/}}

\begin{document}

\begin{titlepage}
\vspace*{-3cm} 
\begin{flushright}
{\tt DESY-26-144}\\
\end{flushright}
\begin{center}
\vspace{1.2cm}

{\LARGE\bfseries One ring to rule them all:\\ syzygy maximization for Argyres--Douglas}

\vspace{1.2cm}
{\large
Monica Jinwoo Kang,$^{1}$ Craig Lawrie,$^{2,3}$ and Jaewon Song$^{4}$\\}
\vspace{.7cm}
{$^1$ Mitchell Institute for Fundamental Physics and Astronomy, Texas A\&M University,\\
College Station, TX 77843, U.S.A.}\par
\vspace{.2cm}
{$^2$ II.~Institut f\"ur Theoretische Physik, Universit\"at Hamburg, 22607 Hamburg, Germany}\par
\vspace{.2cm}
{$^3$ Deutsches Elektronen-Synchrotron DESY,
Notkestr.~85, 22607 Hamburg, Germany}\par
\vspace{.2cm}
{$^4$ Department of Physics, Korea Advanced Institute of Science and Technology\\
Daejeon 34141, Republic of Korea}\par
\vspace{.2cm}

\vspace{.3cm}

\scalebox{.9}{\tt monicak@tamu.edu, craig.lawrie1729@gmail.com, jaewon.song@kaist.ac.kr}\par
\vspace{1.2cm}
\textbf{Abstract}
\end{center}
\noindent We propose a uniform construction of affine schemes encoding the Higgs branches and Macdonald indices of the Argyres--Douglas theories $\mathcal{D}_p(ADE)$, with $\gcd(p,h^\vee)=1$. The construction captures conjectural OPE decoupling relations via the Jacobian ring of a single invariant polynomial potential, obtained from a polynomial twisted period of the corresponding Coxeter orbit space. We show that these potentials are ``syzygy-maximizing'' in the first jet schemes and recover the Higgs branches as their reduced critical loci. We obtain closed determinant expressions for the classical types and explicit potentials for $E_6$, $E_7$, and $E_8$. Powers of the nilradical define an intrinsic filtration on the resulting Jacobian rings, from which we compute candidate Macdonald indices with the correct Schur and Hall--Littlewood limits, in the examples studied. Restricting the potentials to Slodowy slices extends the prescription to nilpotently Higgsed theories. We further conjecture that these Jacobian rings are isomorphic to Zhu's $C_2$-algebras of simple affine VOAs at boundary admissible levels and their quantum Drinfeld--Sokolov reductions.
\vspace{0.65cm}
\small
\begin{center}
\includegraphics[height=1.1cm,trim={0 0 0 0},clip]{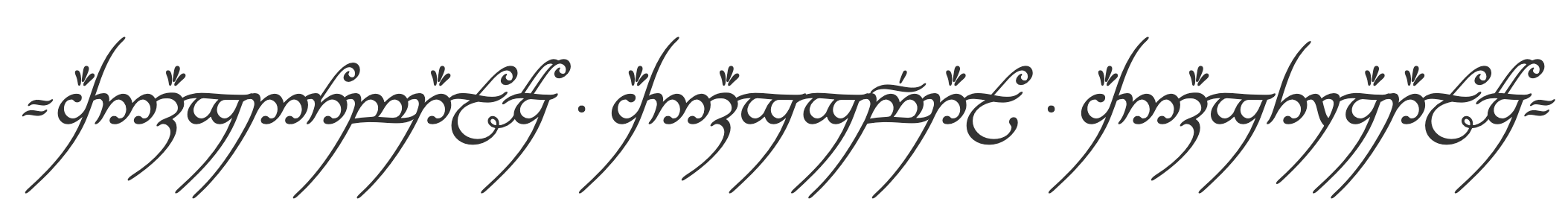}
\hspace{20pt}

\end{center}
\normalsize
\vfill 
\end{titlepage}


\tableofcontents
\newpage

\section{Introduction} \label{sec:intro}

It has been a long-standing problem to specify quantum field theory in a fully non-perturbative way, especially since the discovery of the existence of a rich landscape of non-Lagrangian field theories \cite{Argyres:1995jj, Argyres:1995xn, Minahan:1996fg, Minahan:1996cj, Gaiotto:2009we}. Even though this may be a dream that is far-fetched for a full solution, we can at least hope for finding a `solvable sub-sector' for a supersymmetric theory. 
In fact, supersymmetry makes part of this problem accessible: protected operators and their operator product expansions can often be studied exactly, even at strong coupling. In four-dimensional $\mathcal N=2$ superconformal field theories, these operators give rise to several closely related structures, including the Higgs branch chiral ring, the Schur and Macdonald indices \cite{Gadde:2011uv}. Understanding how these structures fit together offers a way to study theories for which a direct description of the full dynamics is unavailable.

The Higgs branch is a natural starting point, but its geometry captures only part of the protected operator content. A theory whose Higgs branch is a point can still have a nontrivial Schur sector. In a recent series of papers \cite{Kang:2025zub, Kang:2026nge, Kang:2026ona, SUPERPOINT}, we proposed that, for the 4d $\mathcal{N}=2$ superconformal theories, their $1/4$-BPS protected spectrum can be succinctly encoded in a bifiltered affine scheme; in turn, additional information is retained by a commutative ring with nilpotent elements, equipped with a grading and a compatible filtration. This geometric object can be systematically found by postulating a set of `generators' and `relations', which are translated to the key supermultiplets and their decoupling structure of the operator product expansions (OPE). Its reduced spectrum is the Higgs branch, while the full ring records proposed relations among protected operators. Including their derivatives then gives a prescription for the Schur and Macdonald indices. The central question is how to determine this ring from physical data.

In this paper, we give a uniform construction for the Argyres--Douglas theories $\mathcal{D}_p(G)$, with $G$ of $ADE$ type, $p\geq2$, and $\gcd(p,h^\vee)=1$, where $h^\vee$ is the dual Coxeter number \cite{Cecotti:2012jx, Cecotti:2013lda}. These theories provide a large family in which the Higgs branches and Schur indices are known \cite{Buican:2015ina, Cordova:2015nma, Song:2015wta, Song:2017oew}, while general Macdonald indices remain difficult to obtain \cite{Song:2016yfd, Buican:2015tda, Song:2017oew, Agarwal:2018zqi, Foda:2019guo, Maruyoshi:2016aim, Agarwal:2016pjo, Maruyoshi:2016tqk}. We construct a single invariant polynomial potential for each theory, whose derivatives define the proposed ring. The construction gives closed determinant expressions for types $A$ and $D$, and explicit potentials for $E_6$, $E_7$, and $E_8$, with a recursive prescription for all allowed $p$. The same potential also determines our proposal after nilpotent Higgsing.

Let us first explain why the nonreduced ring is useful. 
For example, it is known that the Argyres--Douglas theory of type $(A_1, A_{2n})$ has a schematic vanishing relation of the form $T^{n+1} \sim 0$, where $T$ is the stress-tensor multiplet \cite{Liendo:2015ofa, Agarwal:2018zqi}. More precisely, this means that a short multiplet that appears on the right-hand side generically under the OPE is absent. This motivates us to write a ring defined by $R= \mathbb{C}[x]/(x^{n+1})$: 
\begin{equation}
  \begin{aligned}
    \left(\widehat{\mathcal{C}}_{0(0,0)}\right)^{n+1} \,\not\supseteq\, \widehat{\mathcal{C}}_{n(\tfrac{n}{2},\tfrac{n}{2})} \quad &\leadsto \quad x^{n+1} = 0 \\
    &\leadsto \quad \mathbb{C}[x] \,\, / \, \left( x^{n+1} \right) \,,
  \end{aligned}
\end{equation}
where we have used the standard superconformal multiplet notation of \cite{Dolan:2002zh}.
Here $x$ represents the stress-tensor class in the proposed commutative description and has $(q,T)$-degree $(2,1)$.\footnote{For this particular example with only one variable, we do not actually need to introduce two separate degrees. For multi-variables, this is crucial.} Although the reduced spectrum is a single point for every $n$, the rings are different: they retain $x,x^2,\ldots,x^n$. The nilpotency relation therefore records information that is lost upon passing to the Higgs branch. More generally, our proposal associates to a theory an affine scheme $\operatorname{Spec}(R)$, which is the spectrum of this polynomial ring quotient. The term ``nonreduced'' means that $R$ contains nonzero nilpotent elements: elements whose sufficiently high powers vanish. In the theories considered here, $R$ is commutative: theories containing fermionic generators are discussed in \cite{SUPERPOINT}.

The next step is to include derivatives. If $R=\mathbb{C}[x_a]/(f_\alpha)$, introduce formal series
\begin{equation}
    x_a (t)=\sum_{j\geq 0}x_{a,j}t^j \,,
\end{equation}
and impose every coefficient of $f_\alpha(x(t))=0$. The resulting ring, denoted $J_\infty (R)$, is the coordinate ring of the arc space of $\operatorname{Spec}(R)$.\footnote{An arc space $J_\infty(X)$ of a variety or scheme $X$ is a collection of formal paths (arcs) in $X$.} Keeping coordinates and equations only through order $t^m$ gives the coordinate ring $J_m(R)$ of the $m$-th jet scheme. In the proposed operator interpretation, the new coordinates account for derivatives of the generators, and the coefficient equations account for derivatives of their relations. This makes the arc space a natural object for counting the protected operators together with their descendants.\footnote{When $X$ is smooth, $J_1(X)$ is the total space of its tangent bundle. More generally, it parametrizes tangent vectors to the scheme $X$.
It turns out that the Hilbert series (in an appropriate limit that we will discuss soon) of $J_\infty(R)$ captures the Schur and Macdonald indices.}

Let us now state the index proposal construction. We equip $R$ with a $q$-grading and a compatible increasing, multiplicative $T$-filtration. Here, $F_sR$ consists of elements of filtration degrees at most $s$, and multiplication maps $F_sR\cdot F_tR$ into $F_{s+t}R$. We denote the underlying graded ring, with the filtration forgotten, by $R_S$. In our convention, the index proposal provides the Macdonald index to be
\begin{equation}
    \mathcal I_{\mathrm{Mac}}(q,T)=\operatorname{Tr}_{\mathcal{H}_{\mathrm{Mac}}} (-1)^F q^{\Delta-R}T^{R-r} \,,
\end{equation}
where $\Delta$ is the conformal dimension and $R,r$ are the $SU(2)_R$ and $U(1)_r$ charges, respectively. The trace is taken over the protected sector obeying $\Delta-2j_1-2R-r=0$ and $j_1-j_2+r=0$, with $j_1,j_2$ the Lorentz spins \cite{Dolan:2002zh,Gadde:2011uv}. Our index proposal is to compute the appropriate Hilbert Series
\begin{equation}
    \label{eq:indexproposal}
    \mathcal{I}_{\mathrm{Mac}}(q,T) =\operatorname{HS}_{q,q,T}\!\left(\operatorname{gr} J_\infty(R)\right) \,,\qquad
    \mathcal{I}_{\mathrm{Schur}}(q)=\operatorname{HS}_{q,q}\!\left(J_\infty(R_S)\right) \,.
\end{equation}
The two occurrences of $q$ identify the intrinsic grading and jet order: taking one derivative increases the total $q$-degree by one. Jet derivatives preserve the filtration, and $T$ records the associated graded degree. The associated graded ring is then taken after forming the coordinate ring of the arc space. Setting $T=1$ forgets this refinement and gives the Schur index. See \cite{Andrews:2025krn,Bhargava:2023hsc} for the essentially same construction as ours for the $(A_1,A_{2n})$ and $(A_1,D_{2n+1})$ families.

The filtered graded affine scheme $\operatorname{Spec}(R)$ is proposed to be constructed from OPE decoupling phenomena in the $1/4$-BPS sector of the 4d theory. Such decoupling has been studied in certain Argyres--Douglas theories in \cite{Agarwal:2018zqi,Liendo:2015ofa, Song:2021dhu}. However, for general 4d $\mathcal{N}=2$ SCFTs, it can be technically challenging to discover such phenomena. Using the SCFT/VOA correspondence \cite{Beem:2013sza}, the OPE decoupling can be mapped to the question of finding null relations in the VOA; however, this is also often computationally expensive, and relies on the correct identification of the VOA. On the other hand, a strategy to produce a candidate $R_S$, the $q$-graded ring ignoring the $T$-filtration, was put forth in \cite{Kang:2026nge,Kang:2026ona}. Under our proposal, the Schur index is
\begin{equation}
    I_\text{Schur}(q) = \lim_{T \rightarrow 1} I_\text{Mac}(q,T) = \lim_{T \rightarrow 1} \operatorname{HS}_{p,q,T}(\operatorname{gr} J_\infty(R)) \Big|_{p \rightarrow q} = \operatorname{HS}_{p,q}(J_\infty(R_S)) \Big|_{p \rightarrow q} \,.
\end{equation}

In \cite{Kang:2026nge,Kang:2026ona}, we used this observation to constrain candidate rings. Matching the known Schur index, together with the expected Higgs branch, requires particular minimal syzygies in the first jet scheme. Here ``minimal'' means that the syzygy cannot be obtained from lower-degree ones by multiplication with coordinate polynomials. Requiring such an additional relation restricts the coefficients of the defining equations. We call this procedure \emph{syzygy maximization}, since the corresponding algebraic Betti number increases at special values of those coefficients. In the examples studied previously, it often fixes the defining ideal. The present work identifies a uniform construction behind these calculations and also examines when further input is needed.
The resulting rings, equipped with a proposed $T$-filtration, reproduce the Macdonald index where it is known, or its known limits otherwise. 

For $\mathcal{D}_p(SU(N))$, the rings found in \cite{Kang:2026ona} admit a particularly simple presentation: all defining equations are derivatives of a flavor-invariant polynomial.
In this paper, we give a uniform construction of such potentials, together with the generalization to the $\mathcal{D}_p(G)$ SCFTs,\footnote{See \cite{Bonelli:2011aa,Cecotti:2012jx,Xie:2012hs,Cecotti:2013lda,Wang:2015mra,Wang:2018gvb} for the foundational works on this class of theories, or \cite{Couzens:2023kyf} for a recent review.} with $G$ of ADE type and $\gcd(p,h^\vee)=1$. 
Let $\mathfrak{g}_{\mathbb{C}}$ be the complexified Lie algebra of $G$, with adjoint group action by $G_{\mathbb{C}}$. We construct a homogeneous invariant polynomial $P$ of degree $p+1$ and define
\begin{equation}
    R_S = (R_P)_S
    = \mathbb{C}[\mathfrak{g}_{\mathbb{C}}]/J_P \,,
    \qquad
    J_P = (\partial_{x_1}P,\ldots,
           \partial_{x_{\dim\mathfrak{g}}}P) \,,
\end{equation}
where the $x_a$ (for $a=1, \cdots, \dim\mathfrak{g}$) are linear coordinates on $\mathfrak{g}_{\mathbb{C}}$, each of $q$-degree one. $J_P$ is the Jacobian ideal generated by derivatives. 
Thus, $P$ packages all the proposed commutative relations into a single polynomial. We establish that its reduced critical locus is the Higgs branch and that its first jet scheme has the required minimal singlet syzygies.

The construction of $P$ takes place first on a Cartan subalgebra $\mathfrak{h}$, where invariant polynomials become Weyl-invariant polynomials. We relate the equation governing the distinguished syzygy to \emph{polynomial twisted periods} of the Coxeter orbit space $\mathfrak h/W$ \cite{MR2070050,MR2999308}. These are Weyl-invariant polynomials satisfying a differential equation determined by the root system. A polynomial twisted period of degree $p+1$ and parameter $\nu=p/h^\vee$ has a unique invariant extension to $\mathfrak{g}_{\mathbb{C}}$ by Chevalley restriction; this extension is our potential $P$:
\begin{equation}\label{eqn:intro-residue}
    P(X) \propto
    \begin{cases}
        \displaystyle
        \operatorname*{Res}_{z=\infty}
        \det(z\mathbf{1}_N-X)^{p/N}\,dz \,,
        & \mathfrak{g}_{\mathbb{C}}=\mathfrak{sl}_N(\mathbb{C}) \,,
        \\[8pt]
        \displaystyle
        \operatorname*{Res}_{z=\infty}
        z^{-p/(N-1)}
        \det(z\mathbf{1}_{2N}-X)^{p/(2N-2)}\,dz \,,
        & \mathfrak{g}_{\mathbb{C}}=\mathfrak{so}_{2N}(\mathbb{C}) \,.
    \end{cases}
\end{equation}
Here the determinants are taken in the defining representations, and the fractional powers are expanded formally at infinity, with the leading term $z^p$, and the residue extracts minus the coefficient of $z^{-1}$.
Overall normalization of $P$ does not affect its Jacobian ideal $J_P$. These expressions determine the potentials for arbitrary rank and every allowed $p$, without solving separately for their coefficients. 

For the exceptional groups, polynomial twisted periods can instead be constructed recursively from the Saito flat basic invariants \cite{MR2999308,MR558611}. Each step increases $p$ by $h^\vee$. The initial calculation involves only six, seven, or eight Cartan coordinates, after which Chevalley restriction determines the invariant polynomial on the full Lie algebra. This avoids a separate coefficient search in $78$, $133$, or $248$ adjoint coordinates. Table \ref{tbl:potentials} gives explicit exceptional potentials in a trace basis, using the available constructions of flat invariants~\cite{Abriani,Talamini:2010jg}. The derivatives defining the Jacobian ideal are still taken in the full adjoint coordinates. Thus the exceptional theories enter the same construction as the classical families, with explicit equations from which the proposed protected rings can be studied.

There is an important qualification concerning uniqueness. When the next-to-minimal invariant degree is even, the Higgs branch and the required syzygy can allow inequivalent Jacobian rings. The polynomial twisted period gives a distinguished choice, corresponding to the prescribed Saito flat invariant in the syzygy equation. In the $\mathcal{D}_3(SO(10))$ and $\mathcal{D}_5(SO(10))$ examples, explicit VOA null relations select this choice precisely. We discuss this additional input in Section \ref{sec:nonunique}.

The Macdonald refinement requires one further ingredient: the $T$-filtration. We propose that it is determined intrinsically by the nilradical $\mathfrak n=\sqrt{0}\subset(R_P)_S$, the ideal of all nilpotent elements. For the degree-$d$ component, set\footnote{Recall that if an ideal $I$ is generated by $\{f_a\}$, then $I^n$ is generated by all products $f_{a_1}\cdots f_{a_n}$, with repetitions allowed.}
\begin{equation}
    F_s(R_P)_d
    =
    \left(\mathfrak{n}^{\max\{\lceil d-s\rceil,0\}}\right)_d \,,
    \qquad \mathfrak{n}^0=(R_P)_S \,.
\end{equation}
This prescription relates the additional information retained by the $T$-filtration to the nonreduced structure of the scheme. In particular, the diagonal component of the associated graded ring is the Higgs branch chiral ring, so that the Hall--Littlewood limit of the proposed arc space Hilbert series recovers the Higgs branch Hilbert series. Together with the potential $P$, this specifies our candidate filtered graded ring for $\mathcal{D}_p(G)$.

The proposed relation between the filtered graded ring and OPE decoupling involving protected operators also suggests a connection to the vertex operator algebra (VOA) associated to the 4d theory \cite{Beem:2013sza}. In particular, our construction leads to conjectural presentations of Zhu's $C_2$-algebras for the theories considered here. For a VOA $\mathbb{V}$, its $C_2$-algebra is the graded commutative Poisson algebra \cite{MR1317233}
\begin{equation}
    R_{\mathbb{V}}=\mathbb{V}/C_2(\mathbb{V}) \,,
    \qquad
    C_2(\mathbb{V})
    =\operatorname{Span}
    \left(a_{(-h_a-1)}b \,\,|\,\, a,b\in\mathbb{V} \right) \,,
\end{equation}
where $h_a$ is the conformal weight of $a$, and we use the physics mode convention $Y(a,z)=\sum_n a_{(n)}z^{-n-h_a}$. Under the SCFT/VOA correspondence \cite{Beem:2013sza}, the VOA conjecturally associated to $\mathcal{D}_p(G)$ is the simple affine VOA $L_k(\mathfrak{g}_{\mathbb{C}})$ at level $k=-h^\vee+h^\vee/p$ \cite{Xie:2016evu,Song:2017oew}. We conjecture that
\begin{equation}\label{eq:intro-C2}
    R_{L_k(\mathfrak{g}_{\mathbb{C}})}
    \simeq (R_P)_S
    =\mathbb{C}[\mathfrak{g}_{\mathbb{C}}]/J_P \,,
\end{equation}
as graded commutative rings. This identifies the full $C_2$-algebra, including its nilpotent elements, and is therefore stronger than the identification of its reduced spectrum, Arakawa's associated variety \cite{MR3456698}, with the Higgs branch \cite{Beem:2017ooy,Song:2017oew, Xie:2019zlb}.
In fact, it was proposed earlier that the non-reduced associated scheme, given by $X_{\mathbb{V}} = \operatorname{Spec}(R_{\mathbb{V}})$, is an absolute invariant of 4d $\mathcal{N}=2$ SCFT \cite{RastelliTalk1, RastelliTalk2}. Indeed, our result can be thought of as further evidence of this conjecture. However, we would like to stress that not only is $R_P$ a refined version of $R_{\mathbb{V}}$ via $T$-filtration, but it is also, in general, not necessarily identical \cite{SUPERPOINT}.  

The identification of the $C_2$-algebra does not by itself imply our index proposal. The Li filtration on a VOA \cite{MR2172688} gives a canonical surjection of differential algebras \cite{MR2048777,MR3456698,MR3971726}
\begin{equation}
    J_\infty(R_{\mathbb{V}})
    \twoheadrightarrow \operatorname{gr}_{\mathrm{Li}}\mathbb{V} \,.
\end{equation}
Recovering the Schur index from the arc algebra requires this map to be an isomorphism, a property known as classical freeness
\cite{MR4321999,MR4779563}. We conjecture this property for the VOAs considered here.\footnote{More generally, classical freeness has been conjectured to hold for all VOAs relevant for the SCFT/VOA correspondence in \cite{RastelliTalk1,RastelliTalk2}. It has been recently proved for the $\mathfrak{sl}(2)$ affine vertex algebra at the boundary admissible level \cite{arakawa2026clfrsl2}, which corresponds to $\mathcal{D}_p (SU(2))$ for odd $p$. We provide some counterexamples to this conjecture in \cite{SUPERPOINT}.} The Macdonald refinement further requires the proposed $T$-filtration to reproduce the filtration associated with the four-dimensional R-symmetry charges \cite{Song:2016yfd,Beem:2019tfp}.

The same potential also supplies a proposal for the theories $\mathcal{D}_p(G,O)$ obtained by nilpotent Higgsing. Choose an $\mathfrak{sl}_2$-triple $(e,h,f)$ with $e\in O$, and let $\mathcal{S}_e=e+\ker(\operatorname{ad}f)$ be the corresponding Slodowy slice. Following the prescription of \cite{Kang:2026ona}, we restrict the potential to the slice, $P_e=P|_{\mathcal{S}_e}$, before taking its derivatives. For the associated simple $W$-algebra, obtained by Drinfeld--Sokolov reduction \cite{Beem:2014rza,Xie:2016evu,Song:2017oew}, we conjecture
\begin{equation}\label{eq:intro-W}
    R_{\mathcal{W}_k(\mathfrak{g}_{\mathbb{C}},f)}
    \simeq \mathbb{C}[\mathcal{S}_e]/J_{P_e} \,,
\end{equation}
where the grading on the right-hand side is the half-Kazhdan grading. Thus, restriction of the potential to a Slodowy slice provides a proposed realization of nilpotent Higgsing at the level of the full $C_2$-algebra.

We also discuss a connection to the spectral geometry of these theories. In the type-$A$ examples, the polynomial entering our residue formula also defines a deformation subfamily of the Seiberg--Witten curve. Together with our conjectural identification of the resulting Jacobian ring with Zhu's $C_2$-algebra, this suggests a route from Coulomb branch spectral data to the associated VOA, at the level of its commutative $C_2$-algebra. For the exceptional groups, the adjoint quotient and its cameral cover provide a framework for comparing the invariant data entering our potentials with those of the Hitchin system, although a geometric explanation for the selection of the particular polynomial twisted period remains open.

The organization of this paper is as follows: In Section \ref{sec:unip}, we derive the universal potential for $\mathcal{D}_p(G)$ Argyres--Douglas theories, which completely fixes $(R_P)_S$. In Section \ref{sec:Punique}, we show that the syzygy we find is indeed minimal, so that it indeed solves the syzygy-maximization. We also discuss its uniqueness.
In Section \ref{sec:Macdonald}, we introduce the $T$-filtration for $\mathcal{D}_p(G)$ theories and compute Macdonald indices. In Section \ref{sec:slod}, we discuss the nilpotent Higgsing realized as a Slodowy slice restriction. In Section \ref{sec:Hitchin}, we make a connection of our universal potential to the spectral curve for the Hitchin system or equivalently to the Seiberg--Witten curve. Then we conclude with discussions in Section \ref{sec:discussion}. We also provide an interactive website ArcsToSCFT (\url{https://jws611.github.io/ArcsToSCFT-atlas}), to navigate through indices for a large number of $\mathcal{D}_p(G, \mathcal{O})$ theories. 

\section{The universal potential for \texorpdfstring{\boldmath{$\mathcal{D}_p(G)$}}{Dp(G)}}\label{sec:unip}

We consider the $\mathcal{D}_p(G)$ family of theories, where $p \geq 2$ is an integer and $G$ is a compact simple Lie group of ADE type. Let $\mathfrak{g}$ denote the simple and simply-laced Lie algebra associated to $G$, and let $\mathfrak{g}_\mathbb{C}$ be its complexification. We let $G_\mathbb{C}$ denote the simple and simply-connected complex Lie group with Lie algebra $\mathfrak{g}_\mathbb{C}$. Furthermore, we introduce an invariant nondegenerate bilinear form on $\mathfrak{g}_\mathbb{C}$, $(-, -)$, which is proportional to the Killing form and unique up to overall normalization. Throughout this section, we assume that $\mathfrak{g} \neq \mathfrak{su}(2)$ and that $\gcd(p, h^\vee) = 1$, where $h^\vee$ is the dual Coxeter number of $\mathfrak{g}$. The latter condition ensures that the Higgs branch, $\mathcal{M}_\mathcal{H}^{(p,G)}$, is the closure of a $\mathfrak{g}_\mathbb{C}$ nilpotent orbit.

Treating $\mathfrak{g}_\mathbb{C}$ as an affine space, its coordinate ring is isomorphic to the polynomial ring in $d = \operatorname{dim}(\mathbf{adj})$ indeterminates, where $\mathbf{adj}$ is the adjoint representation of $\mathfrak{g}_\mathbb{C}$:
\begin{equation}\label{eqn:CgCpoly}
    \mathbb{C}[\mathfrak{g}_\mathbb{C}] \cong \mathbb{C}[x_1, \cdots, x_d] \,.
\end{equation}
We consider this as a graded ring where the grading is specified by the degree assignment: $\operatorname{deg}(x_a) = 1$.
We consider the subring of $\mathbb{C}[\mathfrak{g}_\mathbb{C}]$ which consists of $G_\mathbb{C}$-invariant polynomials; it is well-known that this is also isomorphic to a graded polynomial ring:
\begin{equation}
    \mathbb{C}[\mathfrak{g}_\mathbb{C}]^{G_\mathbb{C}} \cong \mathbb{C}[u_1, \cdots, u_r] \,, 
\end{equation}
where $r$ is the rank of $\mathfrak{g}_\mathbb{C}$, and the $u_i$ have degrees $d_i$ which are the degrees of the fundamental Casimir invariants of $\mathfrak{g}_\mathbb{C}$.

To determine the graded ring underlying both the Schur index and the Higgs branch of $\mathcal{D}_p(G)$, we select a degree $(p+1)$ $G_\mathbb{C}$-invariant potential, $P$:
\begin{equation}
    P \in \mathbb{C}[\mathfrak{g}_\mathbb{C}]^{G_\mathbb{C}} \,.
\end{equation}
From $P$, we consider the Jacobian ideal $J_P$ and the corresponding quotient ring $R_P$:
\begin{equation}
    J_P = (\partial_{x_1} P, \cdots, \partial_{x_d} P) \,, \qquad R_P = \mathbb{C}[\mathfrak{g}_\mathbb{C}] \, / \, J_P \,.
\end{equation}
We seek a $P$ such that the reduced spectrum of $R_P$ is the Higgs branch:
\begin{equation}\label{eqn:BIGHIGGS}
    \operatorname{Spec}(R_P)_\text{red} = \mathcal{M}_\mathcal{H}^{(p,G)} \,.
\end{equation}
Since we are dealing with Jacobian rings, we often use the identification between $\operatorname{Spec}(R_P)_\text{red}$ and the critical points of $P$, denoted $\operatorname{Crit}(P)$.
For $p > h^\vee$ the Higgs branch is the nilpotent cone of $\mathfrak{g}_\mathbb{C}$ (referred to as $\mathcal{N}$), whereas for $p < h^\vee$, it is given by the closure of some non-maximal nilpotent orbit of $\mathfrak{g}_\mathbb{C}$. The Higgs branch is proposed to be the associated variety \cite{MR3456698} of the simple affine vertex operator algebra $L_k(\mathfrak{g}_\mathbb{C})$ at boundary admissible level $k = -h^\vee + h^\vee/p$:\footnote{This also agrees with the maximal allowed partitions as observed from the consistency of supergravity solutions in \cite{Couzens:2023kyf}.}
\begin{equation}\label{eqn:HiggsAD}
    \mathcal{M}_\mathcal{H}^{(p,G)} = \overline{\mathcal{O}}_p = \left\{ X \in \mathfrak{g}_\mathbb{C} \,|\, (\operatorname{ad}_X)^{2p} = 0 \right\} \,.
\end{equation}
Furthermore, we require that $P$ is such that the algebraic Betti number of the first jet scheme satisfies:
\begin{equation}\label{eqn:BIGSYZYGY}
    \beta_{2,(p+c-1, 1)}(J_1(R_P)) \big|_\text{singlet} = \begin{cases}
        2 \quad \text{ for } \mathfrak{g} = \mathfrak{so}(8) \,, \\    
        1 \quad \text{ otherwise } \,.
    \end{cases}
\end{equation}
Here $c$ is the next-to-minimal fundamental Casimir degree of $\mathfrak{g}_\mathbb{C}$.\footnote{The $\mathfrak{so}(8)$ case is exceptional because there are \emph{two} independent fundamental Casimir invariants of the next-to-minimal degree. The $\mathfrak{su}(2)$ case is also exceptional, since there is no next-to-minimal fundamental Casimir, and any invariant degree $p+1$ polynomial is necessarily $P(X) \propto C_2(X)^{\frac{p+1}{2}}$. } In \cite{Kang:2026nge,Kang:2026ona}, for $\mathfrak{g} = \mathfrak{su}(N)$, this Betti number condition was observed from matching the closed-form expression for the Schur index with the Hilbert series of the arc space written in terms of the algebraic Betti numbers. For $\mathcal{D}_p(G)$ with $\gcd(p, h^\vee) = 1$, the Schur index is also known \cite{Song:2015wta,Song:2017oew,Xie:2016evu,kacwakimoto}:
\begin{equation}\label{eqn:Schurgeneral}
    I_{\mathcal{D}_p(G)}(q, \vec{z}) = \operatorname{PE}\left[\frac{q-q^p}{(1-q)(1-q^p)} \chi_\textbf{adj}(\vec{z}) \right]\,,
\end{equation}
where $\operatorname{PE}$ is the plethystic exponential, and $\chi_\textbf{adj}(\vec{z})$ is the character of the adjoint representation of the $\mathfrak{g}$ flavor symmetry. The same analysis for general $G$ then leads to equation \eqref{eqn:BIGSYZYGY}. In the remainder of this section, we would like to find a closed-form expression or a construction for such a $P$, and to explore the extent to which $P$ is unique.

\subsection{An equation for the singlet syzygy}\label{sec:syzygyeqn}

In this subsection, we derive the equation for a singlet syzygy as in equation \eqref{eqn:BIGSYZYGY} to exist. 
We consider $X \in \mathfrak{g}_\mathbb{C}$, and we write the first jet expansion as $X = X_0 + X_1 t$, where $X_0, X_1 \in \mathfrak{g}_\mathbb{C}$.
Write $F = \nabla P$, where the gradient is taken with respect to the bilinear form $\langle -,- \rangle$. The relations in the first jet scheme of $\operatorname{Spec}(R_P)$ are defined by the coefficients of $t^0$ and $t$ in $F(X_0 + X_1 t) = F_0 + F_1 t + \cdots$. Thus, we find
\begin{equation}\label{eq:F0F1}
    F_0 = \nabla P (X_0) \,, \qquad F_1 = H_P(X_0)X_1 \,,
\end{equation}
where $H$ is the usual Hessian:
\begin{equation}
    H_f(X)Y = \left.\frac{d}{ds}\nabla f(X+sY)\right|_{s=0} \,.
\end{equation}
Here $f \in \mathbb{C}[\mathfrak{g}_\mathbb{C}]$ and $X, Y \in \mathfrak{g}_\mathbb{C}$. Passing to coordinates via the isomorphism in equation \eqref{eqn:CgCpoly}, let $S_1=\mathbb{C}[x_{a,0},x_{a,1}]$, $\deg(x_{a,0})=(1,0)$, and $\deg(x_{a,1})=(1,1)$. The first jet scheme of $\operatorname{Spec}R_P$ has coordinate ring $J_1(R_P)$, obtained by quotienting $S_1$ by the ideal generated by the components of $F_0$ and $F_1$ in equation \eqref{eq:F0F1}. All algebraic Betti numbers below refer to its minimal bigraded free resolution over $S_1$. 
Let $\mathcal{K}$ denote the module of relations among these defining equations, and let $\mathfrak{m}_1=(x_{a,0},x_{a,1})$ be the homogeneous maximal ideal of \(S_1\). Then the second Betti number of bidegree $(b, c)$ in the $G$-singlet is given as
\begin{equation}\label{eq:betti-quotient}
	\beta^{S_1}_{2,(b,c)}(J_1(R_P))_{\mathrm{singlet}} = \dim\bigl((\mathcal{K}/\mathfrak m_1\mathcal{K})_{(b,c)}^{G_{\mathbb{C}}}\bigr) \ , 
\end{equation}
where we evaluate the dimension of $G_\mathbb{C}$-invariant bidegree $(b, c)$ part of the quotient. 
The quotient removes relations obtained by multiplying lower-degree syzygies by coordinate polynomials. Thus a syzygy contributes to the required Betti number only if its coefficient vector represents a nonzero class in this quotient.

In equation \eqref{eqn:BIGSYZYGY}, we proposed the necessary minimal singlet syzygies for the Hilbert series of the arc space to match the Schur index of $\mathcal{D}_p(G)$. A singlet syzygy admits a representative of the form
\begin{equation}
    \langle A(X_0, X_1), F_0 \rangle + \langle B(X_0, X_1), F_1 \rangle = 0 \,,
\end{equation}
where $A$ and $B$ are polynomials in the adjoint representation of $G_\mathbb{C}$. Recall that we assign bidegrees $(1,0)$ and $(1,1)$ to $X_0$ and $X_1$. Since $P$ has degree $p+1$, $F_0$ and $F_1$ have bidegrees $(p,0)$ and $(p,1)$, respectively.
Consequently, a syzygy of bidegree $(p+c-1,1)$ requires 
\begin{equation}
    \operatorname{deg}(A) = (c-1, 1) \,, \qquad \operatorname{deg}(B) = (c-1, 0) \,.
\end{equation}
Thus $A$ is linear in $X_1$ and homogeneous of degree $c-2$ in $X_0$, whereas $B$ depends only on $X_0$ and is homogeneous of degree $c-1$. Then, we can find bases for the spaces of $G_{\mathbb{C}}$-equivariant polynomial maps $A$ and $B$ at these degrees, and thus express all possible singlet syzygies.

Let us begin by studying the cases where $c$ is odd, which are markedly more straightforward. 
When $c$ is odd, the Casimir polynomial $C_c(X)$ is unique up to normalization, whereas for $c$ even, we can define a new basis of the $G_\mathbb{C}$-invariant polynomial ring via
\begin{equation}
    \widetilde{C}_c(X) = C_c(X) + a C_2(X)^{c/2} \,,
\end{equation}
for arbitrary $a$.
For $\mathfrak{g} = \mathfrak{su}(N)$, there are only three terms which can contribute in a singlet syzygy of degree $(p-1+c, 1)$:
\begin{equation}
    \langle \nabla C_3(X_0), F_1 \rangle \,, \quad  \langle H_{C_3}(X_0)X_1, F_0 \rangle \,, \quad \langle \operatorname{ad}_{X_0}(X_1), F_0 \rangle \,,
\end{equation}
whereas for $\mathfrak{g} = \mathfrak{e}_6$, we have the following four terms:
\begin{equation}
    \langle \nabla C_5(X_0), F_1 \rangle \,, \quad  \langle H_{C_5}(X_0)X_1, F_0 \rangle \,, \quad  \langle C_2(X_0)\operatorname{ad}_{X_0}(X_1), F_0 \rangle \,, \quad \langle \operatorname{ad}_{X_0}^3(X_1), F_0 \rangle \,.
\end{equation}
Here $\operatorname{ad}_X(Y) = [X, Y]$. Since $P$ is a $G_\mathbb{C}$-invariant polynomial, then any term of the form
\begin{equation}\label{eqn:killed}
    \langle C_2(X_0)^k \operatorname{ad}^\ell_{X_0}(Y), F_0 \rangle \,,
\end{equation}
for $k \geq 0$ and $\ell \geq 1$, is identically zero.
More specifically, adjoint invariance of $P$ implies $\operatorname{ad}_{X_0}F_0=0$, whose components are syzygies of bidegree $(p+1,0)$. Every term of the form in equation \eqref{eqn:killed} is a polynomial combination of these lower-degree syzygies and therefore does not contribute a minimal syzygy at bidegree $(p+c-1,1)$.
Thus, any candidate minimal syzygy of the required bidegree takes the form
\begin{equation}\label{eqn:syzygyCODD}
    a \langle \nabla C_c(X_0), F_1 \rangle + b \langle H_{C_c}(X_0)X_1, F_0 \rangle = 0 \,,
\end{equation}
for some $a, b$ defined up to overall normalization. Using the self-adjointness of the Hessian with respect to the bilinear form, we can rewrite this expression as
\begin{equation}
    \langle a H_P(X_0) \nabla C_c(X_0) + bH_{C_c}(X_0)\nabla P(X_0) , X_1 \rangle = 0 \,.
\end{equation}
Since this must hold for all $X_1$ and the bilinear form is non-degenerate, the existence of the syzygy in equation \eqref{eqn:syzygyCODD} implies that
\begin{equation}
    a \,H_P(X) \nabla C_c(X) + b \,H_{C_c}(X)\nabla P(X) = 0 \,,
\end{equation}
for all $X \in \mathfrak{g}_\mathbb{C}$. In particular, we can take the bilinear form of this object with $X$ itself, and using the Euler relation that 
\begin{equation}\label{eqn:Euler}
    H_f(X)X = (\operatorname{deg}(f) - 1) \nabla f(X) \,,
\end{equation}
we then find that 
\begin{equation}
    (ap + (c-1)b) \langle \nabla P (X), \nabla C_c(X) \rangle = 0 \,,
\end{equation}
where the equality follows from equation \eqref{eqn:syzygyCODD}. This equation has two branches:
\begin{equation}\label{eqn:spyro}
     ap + (c-1)b = 0 \quad \text{ or } \quad \langle \nabla P (X), \nabla C_c(X) \rangle = 0 \,.
\end{equation}

In the latter branch of equation \eqref{eqn:spyro}, we have a syzygy at bidegree $(p+c-1,0)$. We now show that this branch violates the Higgs branch condition in equation \eqref{eqn:BIGHIGGS}. Let us first consider the restriction to a Cartan subalgebra
\begin{equation}
    q = P|_\mathfrak{h} \,, \qquad f = C_c|_\mathfrak{h} \,.
\end{equation}
Since $P$ is $G_\mathbb{C}$-invariant, then
\begin{equation}
    \nabla q (x) = \nabla P(x) \,,
\end{equation}
for any $x \in \mathfrak{h}$. Furthermore, $\nabla f \neq 0$, so a non-trivial syzygy $\langle \nabla P, \nabla C_c \rangle = 0$ would imply the existence of the non-trivial syzygy $\langle \nabla q, \nabla f \rangle = 0$ on the Cartan. The Higgs branch condition in equation \eqref{eqn:BIGHIGGS} requires that the critical points of $P$ are nilpotent, and therefore the critical points of $q$ must be nilpotent, but the only nilpotent element of $\mathfrak{h}$ is the origin:
\begin{equation}
    \operatorname{Crit}(q) = \operatorname{Crit}(P) \cap \mathfrak{h} = \{ 0 \} \,.
\end{equation}
Therefore $\operatorname{Spec}(\mathbb{C}[\mathfrak{h}] \, / \, J_q)$ is a complete intersection. The first syzygy module is generated by the Koszul syzygies, which here have total degree $2p$. Except in the case of $p=2$ and $c=3$, we have that $p > c-1$, and therefore the syzygy $(\nabla q, \nabla f) = 0$ is at too low degree to be Koszul. For $p=2$ and $c=3$, then $P \propto C_3$, and one can show directly that $ \langle \nabla C_3 , \nabla C_3 \rangle \neq 0$. Thus, in both cases, the branch $\langle \nabla P,\nabla C_c \rangle =0$ is incompatible with the Higgs branch condition.

Since the $\langle \nabla P (X), \nabla C_c(X) \rangle = 0 $ branch in equation \eqref{eqn:spyro} is thus excluded, the combination of the requirement that $\operatorname{Spec}(R_P)_\text{red}$ reproduces the Higgs branch of $\mathcal{D}_p(G)$ and that the syzygy condition in equation \eqref{eqn:BIGSYZYGY} is satisfied implies that $P$ must solve the following equation for all $X \in \mathfrak{g}_\mathbb{C}$ (recalling that we assume that $c$ is odd):
\begin{equation}\label{eqn:coddsyzygy}
    (c-1) H_P(X) \nabla C_c(X) - pH_{C_c}(X)\nabla P(X) = 0 \,.
\end{equation}

We now turn to the cases where $c$ is even, which have additional subtleties. In addition to the identically vanishing terms in equation \eqref{eqn:killed}, we must account for a syzygy that follows from the homogeneity of $P$. Self-adjointness of the Hessian gives the syzygy
\begin{equation}
    \langle X_0, F_1 \rangle - p \langle X_1, F_0 \rangle = 0 \,,
\end{equation}
which is of bidegree $(p+1,1)$ and exists for all $P$. We define $m = c/2$ and then we can multiply this relation by $C_2(X_0)^{m-1}$ to obtain a (non-minimal) syzygy of bidegree $(p-1+c, 1)$. Here, we have normalized the quadratic invariant such that 
\begin{equation}
    C_2(X) = \frac{1}{2} \langle X, X \rangle \,.
\end{equation}
Therefore, we can write any new (minimal) singlet syzygy in bidegree $(p-1+c, 1)$ as a sum involving the following four terms:
\begin{equation}
  \begin{aligned}
    T_1 &= \langle \nabla C_c(X_0), F_1 \rangle \,, \qquad
    && T_2 = \langle H_{C_c}(X_0)X_1, F_0 \rangle \,,\\ 
    V &= C_2(X_0)^{m-1} \langle X_1, F_0 \rangle \,, \qquad
    && W = C_2(X_0)^{m-2} \langle X_0, X_1 \rangle \langle X_0, F_0 \rangle \,,
  \end{aligned}
\end{equation}
with coefficients which we call $a, b, v, w$, respectively. Similarly to the $c$ odd case, we can collect the linear terms in $X_1$ to obtain a vector equation, and then pair with $X_0$ to find that the existence of a non-trivial syzygy implies that
\begin{equation}
    (a p + (c-1)b) \langle \nabla P(X), \nabla C_c(X) \rangle + (p+1)(v + 2w)C_2^{m-1}(X) P(X) = 0 \,,
\end{equation}
as a polynomial identity in $\mathbb{C}[\mathfrak{g}_\mathbb{C}]^{G_\mathbb{C}}$. For convenience, let us write this as
\begin{equation}
    \alpha S + \beta Q = 0 \,,
\end{equation}
where 
\begin{equation}
    \begin{gathered}
        \alpha = (a p + (c-1)b) \,, \quad \beta = (p+1)(v + 2w) \,, \\ 
        S(X) =  \langle \nabla P(X), \nabla C_c(X) \rangle \,, \quad Q(X) = C_2^{m-1}(X) P(X) \,.
    \end{gathered}
\end{equation}
Since $Q(X)$ is a nonzero polynomial, there are two branches of solutions, either
\begin{equation}
    \alpha = \beta = 0 \quad \text{ or } \quad \alpha \neq 0, \, S(X) = \kappa Q(X) \,,
\end{equation}
for some constant $\kappa$.
In the latter case, we can change basis of $\mathbb{C}[\mathfrak{g}_\mathbb{C}]^{G_\mathbb{C}}$, by redefining 
\begin{equation}
    C_c' = C_c - \frac{\kappa}{m(p+1)}C_2^m \,.
\end{equation}
Then, the condition becomes $\langle \nabla P, \nabla C_c' \rangle = 0$. Any $P$ that satisfies this condition does \emph{not} also satisfy the condition that the quotient ring by the Jacobian ideal of $P$ reproduces the Higgs branch of the $\mathcal{D}_p(G)$ theory, analogously to the odd $c$ case already discussed. Therefore, we need $\alpha = \beta = 0$ if there exists a minimal syzygy in degree $(p-1+c, 1)$ for which the ring has the correct properties. Similarly, we can rule out $a = b = 0$ by the Higgs branch condition, since that would imply that $P$ is proportional to $C_2(X)^{(p+1)/2}$. Altogether, we can write the syzygy as
\begin{equation}\label{eqn:cevengensyzygy}
    (c-1)T_1 - pT_2 + \mu(W - 2 V) = 0 \,,
\end{equation}
and shifting the basis of Casimir invariants of the $G_\mathbb{C}$-invariant polynomial ring via
\begin{equation}
    C_c' = C_c - \frac{\mu}{mp(m-1)} C_2^m \,,
\end{equation}
then we find that $P$ must satisfy the following equation
\begin{equation}
    (c-1)H_P(X)\nabla C_c'(X) - p H_{C_c'}(X)\nabla P(X) = 0 \,.
\end{equation}
Consequently, the choice of $C_c$ need not be fixed in advance: it suffices to find a degree-$(p+1)$ invariant $P$ for which this equation holds for some primitive degree-$c$ representative, and then the first jet scheme of the quotient by the Jacobian ideal has a non-trivial syzygy of the required bidegree.

Putting everything together, we find that, for $\mathfrak{g} \neq \mathfrak{so}(8)$,\footnote{When $\mathfrak{g} = \mathfrak{so}(8)$, there are two syzygies of the form in equation \eqref{eqn:BIGPDE}, one for each of the two independent degree $c = 4$ invariants. These can be represented by a suitable trace quartic, $C_c$, and the Pfaffian, $\widetilde{C}_c$. The triality group, $S_3$, which permutes the three eight-dimensional irreducible representations of $\mathfrak{so}(8)$, acts on these two quartic invariants via the two-dimensional irreducible representation. The corresponding syzygies form a triality doublet. That is, we want to find a $(P, C_c, \widetilde{C}_c)$ such that these two syzygy equations are solved; the discussion of the general solutions in the remainder of this section generalizes to this $\mathfrak{so}(8)$ case.}
requiring the reduced scheme to reproduce the Higgs branch and imposing the existence of the desired (candidate minimal) singlet syzygy, we can find such a graded ring
\begin{equation}
    R_S = R_P = \mathbb{C}[\mathfrak{g}_\mathbb{C}] \, / \, J_P \,,
\end{equation}
by finding a degree $(p+1)$ $G_\mathbb{C}$-invariant potential $P$ satisfying
\begin{equation}\label{eqn:BIGPDE}
    (c-1) H_P(X)\nabla C_c(X) - p H_{C_c}(X) \nabla P(X) = 0 \,,
\end{equation}
for all $X \in \mathfrak{g}_\mathbb{C}$, and where $C_c$ is an appropriate choice of degree $c$ invariant. In particular, if we can find a pair $(P, C_c)$ such that equation \eqref{eqn:BIGPDE} is solved, then we have determined a $P$ such that the coordinate ring of the first jet scheme of $\operatorname{Spec}(\mathbb{C}[\mathfrak{g}_\mathbb{C}] / J_P)$ has a syzygy of the desired bidegree. For even $c$, the choice of primitive invariant $C_c$ is not fixed by these conditions, and different choices can lead to inequivalent potentials $P$, as we discuss in Section \ref{sec:nonunique}.

\subsection{Polynomial twisted periods and \texorpdfstring{$P$}{P}}\label{sec:ptp}

To find solutions to the Hessian identity in equation \eqref{eqn:BIGPDE}, we take a brief detour into the subject of polynomial twisted periods \cite{MR2070050,MR2999308}. We can choose a Cartan subalgebra $\mathfrak{h} \subset \mathfrak{g}_\mathbb{C}$, with Weyl group $W$. We normalize the bilinear form such that roots have length-squared $2$, identifying roots with vectors in $\mathfrak{h}$. By Chevalley restriction there is an isomorphism 
\begin{equation}\label{eqn:isolift}
    \mathbb{C}[\mathfrak{g}_\mathbb{C}]^{G_\mathbb{C}} \cong \mathbb{C}[\mathfrak{h}]^W \,.
\end{equation}
The strategy will be to first construct a polynomial on $\mathfrak{h}$, and then uplift it through this isomorphism to a polynomial on the full Lie algebra.

The definition of a polynomial twisted period is as follows. Let $q$ be a polynomial belonging to $\mathbb{C}[\mathfrak{h}]^W$ and $\nu \in \mathbb{C}$ be an arbitrary complex parameter. If $q$ satisfies
\begin{equation}\label{eqn:ptpdef}
    H_q(x)u = \nu \sum_{\alpha \in \Phi^+} \frac{\langle \alpha, u \rangle \langle \alpha, \nabla q(x)\rangle}{\langle \alpha, x \rangle} \alpha \,,
\end{equation}
for all $u \in \mathfrak{h}$ and $x \in \mathfrak{h}_\text{reg}$, then $q$ is called a polynomial twisted period. Here, $\mathfrak{h}_\text{reg}$ is the regular locus of the Cartan subalgebra. Furthermore, $\Phi^+$ denotes the set of positive roots of $\mathfrak{g}_\mathbb{C}$. For a non-constant homogeneous polynomial twisted period, the constant $\nu$ is in fact fixed as follows. If we take $u = x$, then we can use the fact that
\begin{equation}\label{eq:twisted}
    \sum_{\alpha \in \Phi^+}  \langle \alpha, v \rangle \alpha = h^\vee v \,,
\end{equation}
to simplify equation \eqref{eqn:ptpdef} to
\begin{equation}
    H_q(x)x = \nu h^\vee \nabla q(x) \,.
\end{equation}
Then, recalling that $q$ should be a degree $(p+1)$ polynomial, we can use the Euler relation (as in equation \eqref{eqn:Euler}) to show that we must have
\begin{equation}
    \nu = \frac{p}{h^\vee} \,.
\end{equation}

Explicit nonconstant homogeneous polynomial twisted periods were constructed in \cite{MR2999308}, in the specific cases where the degree of $q$ is
\begin{equation}
    d_i + k h^\vee \,,
\end{equation}
for arbitrary non-negative integer $k$ and where $d_i$ is the degree of any fundamental Casimir invariant of $\mathfrak{g}_\mathbb{C}$. 
In these solutions, the parameter is of the form
\begin{equation}
    \nu = \frac{d_i - 1}{h^\vee} + k \,.
\end{equation}
When $k = 0$, these polynomial twisted periods are precisely the Saito flat basic invariants \cite{MR558611}.
Our assumption that $\gcd(p, h^\vee) = 1$ ensures that there exists some $i$ and $k$ such that
\begin{equation}
    p = d_i - 1 + k h^\vee \,,
\end{equation}
and thus the construction of \cite{MR2999308} supplies a nonzero polynomial 
\begin{equation}
    q_p \in \mathbb{C}[\mathfrak{h}]^W \,,
\end{equation}
such that $\operatorname{deg}(q_p) = p+1$, and we let $P$ denote the lift of $q_p$ to a polynomial on $\mathfrak{g}_\mathbb{C}$ via the isomorphism in equation \eqref{eqn:isolift}. 

To see that the uplift of $q_p$ to $P$ is a polynomial that we are interested in, we proceed as follows. First, we let $t_c$ denote the Saito flat basic invariant of degree $c$ and we assume that the $C_c$ appearing in equation \eqref{eqn:BIGPDE} is the lift of $t_c$ under equation \eqref{eqn:isolift}.\footnote{Recall, we seek a pair $(P, C_c)$ such that equation \eqref{eqn:BIGPDE} is solved. For the (necessarily $c$ even) possibility of $C_c$ not being the lift of $t_c$, see Section \ref{sec:nonunique}.} This is a polynomial twisted period with parameter $(c-1)/h^\vee$. Furthermore, let us define
\begin{equation}\label{eqn:defCalB}
    \mathcal{B}_x(u,v) = \sum_{\alpha \in \Phi^+} \frac{\langle \alpha, u \rangle \langle \alpha, v \rangle}{\langle\alpha, x \rangle} \alpha \,,
\end{equation}
for $u, v \in \mathfrak{h}$ and $x$ in the regular Cartan. This is symmetric: $\mathcal{B}_x(u,v) = \mathcal{B}_x(v,u)$. Since both $q_p$ and $t_c$ are polynomial twisted periods, they satisfy equation \eqref{eqn:ptpdef} individually:
\begin{equation}
    \begin{aligned}
        H_{q_p}(x) \nabla t_c (x) &= \frac{p}{h^\vee} \mathcal{B}_x \langle \nabla t_c(x), \nabla q_p(x) \rangle \,, \\
        H_{t_c}(x) \nabla q_p (x) &= \frac{c-1}{h^\vee} \mathcal{B}_x \langle \nabla q_p(x), \nabla t_c(x) \rangle \,.    
    \end{aligned}
\end{equation}
Thus the symmetry of $\mathcal{B}_x$ immediately implies that
\begin{equation}\label{eqn:swiper}
    (c-1)H_{q_p}(x) \nabla t_c(x) - p H_{t_c}(x) \nabla q_p (x) = 0 \,,
\end{equation}
for all $x \in \mathfrak{h}_\text{reg}$. Using the invariant lifts $P$ and $C_c$, define
the polynomial map
\begin{equation}
    \mathcal{E}(X) = (c-1)H_{P}(X) \nabla C_c(X) - p H_{C_c}(X) \nabla P (X) \,.
\end{equation}
The map $\mathcal{E} : \mathfrak{g}_\mathbb{C} \rightarrow \mathfrak{g}_\mathbb{C}$ is both polynomial and equivariant. It vanishes on the regular Cartan by equation \eqref{eqn:swiper}, and thus on every regular semisimple element, and therefore everywhere by density; see the proof of Theorem 2 in \cite{MR2737780} for details. Therefore, we have proven that 
\begin{equation}
    \mathcal{E}(X) = 0 \quad \text{ for all } X \in \mathfrak{g}_\mathbb{C} \,,
\end{equation}
and thus a polynomial twisted period $q_p$ as given in \cite{MR2999308} fixes a solution of equation \eqref{eqn:BIGPDE} derived from the Higgs branch and syzygy-maximization conditions.

Feigin and Silantyev \cite{MR2999308} provide a uniform recursive construction of these polynomial twisted periods, starting from the Saito flat basic invariants. For each basic invariant degree $d_i$, the construction gives a sequence
\begin{equation}
    p_\ell = d_i - 1 + \ell h^\vee \,, \qquad \ell = 0, 1, 2, \cdots \,,
\end{equation}
and constructs a degree $(p_{\ell+1} + 1)$ polynomial twisted period with twisting parameter $p_{\ell+1} / h^\vee$, starting from the polynomial twisted period associated to $p_\ell$. Each step increases the polynomial degree by $h^\vee$. This construction applies uniformly to all ADE types, however, for types A and D, the resulting polynomials admit explicit residue expressions \cite{MR2999308}. If we regard $X$ as a matrix in the defining representations of either $\mathfrak{sl}_N(\mathbb{C})$ or $\mathfrak{so}_{2N}(\mathbb{C})$, then 
\begin{equation}\label{eqn:det}
\boxed{P(X) \propto [t^{p+1}] \, \operatorname{det}(\bm{1} - t X)^{p/h^\vee} }\,,
\end{equation}
where $[t^{p+1}]$ denotes the coefficient of $t^{p+1}$ in the formal variable $t$, and where we recall that changing the overall normalization of $P$ is considered irrelevant for the Jacobian ideal.

To express the potentials in a simple and explicit form we proceed as follows. Let $X \in \mathfrak{g}_\mathbb{C}$ and let $\bm{R}$ denote the fundamental representation for $\mathfrak{g} = \mathfrak{su}(N)$, the vector representation for $\mathfrak{g} = \mathfrak{so}(2N)$, and the $\bm{27}$, $\bm{56}$ and $\bm{248}$ for $\mathfrak{e}_6$, $\mathfrak{e}_7$, and $\mathfrak{e}_8$, respectively. We can then consider the homogeneous invariant polynomials
\begin{equation}\label{eq:CkasTr}
    C_k(X) = \frac{1}{k} \operatorname{tr}_{\bm{R}} X^k \,.
\end{equation}
In terms of this particular set of invariant polynomials, we have listed the potentials for various values of $p$ in Table \ref{tbl:potentials}. We see that this result indeed agrees with the result of \cite{Kang:2026ona} for $\mathfrak{g} = \mathfrak{su}(N)$. 
The $C_k(X)$ need not be primitive or algebraically independent, but can be reduced to polynomials in the primitive invariants using the appropriate trace identities. In particular, for $\mathfrak{g}_{\mathbb{C}}=\mathfrak{sl}_N(\mathbb{C})$, the Cayley--Hamilton theorem expresses every $C_k$ with $k>N$ as a polynomial in $C_2,\cdots,C_N$. 

\begin{table}[h!]
    \centering
    \renewcommand{\arraystretch}{1.42}
    \begin{threeparttable}
    \begin{tabular}{ccc}
        \toprule
        $\mathfrak{g}$ & $p$ & $P$ \\\midrule
        \multirow{6}{*}{$\mathfrak{su}(N)$} & $2$ & $C_3$ \\ 
         & $3$ & $C_4 - \tfrac{3}{2N}C_2^2$ \\
         & $4$ & $C_5 - \tfrac{4}{N} C_2 C_3$ \\
         & $5$ & $C_6 - \tfrac{5}{N}C_2C_4 - \tfrac{5}{2N}C_3^2 + \tfrac{25}{6N^2}C_2^3$ \\
         & $6$ & $C_7 - \tfrac{6}{N}C_2C_5 - \tfrac{6}{N}C_3C_4 + \tfrac{18}{N^2} C_2^2 C_3$ \\
         & $7$ & $C_8 - \tfrac{7}{N}C_2C_6 - \tfrac{7}{N}C_3C_5 - \tfrac{7}{2N}C_4^2 + \tfrac{49}{2N^2}C_2^2C_4 + \tfrac{49}{2N^2}C_2C_3^2 - \tfrac{343}{24N^3}C_2^4$ \\[5pt]\midrule
        \multirow{3}{*}{$\mathfrak{so}(2N)$} & $3$ & $C_4 - \tfrac{3}{4(N-1)}C_2^2$ \\
         & $5$ & $C_6 - \tfrac{5}{2(N-1)}C_2C_4 + \tfrac{25}{24(N-1)^2}C_2^3$ \\
         & $7$ & $C_8 - \tfrac{7}{2(N-1)}C_2C_6 - \tfrac{7}{4(N-1)}C_4^2 + \tfrac{49}{8(N-1)^2}C_2^2C_4 - \tfrac{343}{192(N-1)^3}C_2^4$ \\[5pt]\midrule
        \multirow{4.5}{*}{$\mathfrak{e}_6$} & $5$ & $C_6 - \tfrac{5}{432}C_2^3$ \\
         & $7$ & $C_8 - \tfrac{7}{16}C_2C_6 + \tfrac{385}{124416}C_2^4$ \\
         & $11$ & $C_{12} - \tfrac{77}{192}C_6^2 - \tfrac{605}{1728}C_2C_5^2 - \tfrac{209}{864}C_2^2C_8 + \tfrac{2959}{41472}C_2^3C_6 - \tfrac{737}{1990656}C_2^6$ \\[2pt]
         & $13$ & $\begin{aligned}C_2C_{12} &+ \tfrac{9}{2}\,C_6C_8 + \tfrac{45}{14}\,C_5C_9 - \tfrac{455}{192}\,C_2C_6^2 \\&- \tfrac{1325}{1728}\,C_2^2C_5^2 - \tfrac{127}{432}\,C_2^3C_8 + \tfrac{8963}{82944}\,C_2^4C_6 - \tfrac{6347}{11943936}\,C_2^7\end{aligned}$ \\[10pt]\midrule
        \multirow{3}{*}{$\mathfrak{e}_7$} & $5$ & $C_6 - \tfrac{35}{11664}C_2^3$ \\
         & $7$ & $C_8 - \tfrac{7}{27}C_2C_6 + \tfrac{2555}{5038848}C_2^4$ \\
         & $11$ & $C_{12} -\tfrac{605}{1512}C_2C_{10} - \tfrac{11}{32}C_6^2 + \tfrac{341}{6048}C_2^2C_8 - \tfrac{649}{559872}C_2^3C_6 - \tfrac{4499}{6530347008}C_2^6$ \\[5pt]\midrule
        \multirow{3}{*}{$\mathfrak{e}_8$} & $7$ & $C_8 - \tfrac{49}{2^7 \cdot 3^3 \cdot 5^5}C_2^4$ \\
         & $11$ & $C_{12} - \tfrac{121}{2^4 \cdot 3^2 \cdot 5^3}C_2^2C_8 + \tfrac{77077}{2^{11} \cdot 3^6 \cdot 5^9}C_2^6$ \\
         & $13$ & $C_{14} -\tfrac{169}{2^2 \cdot 3^2 \cdot 5 \cdot 7}C_2C_{12} + \tfrac{35321}{2^6 \cdot 3^5 \cdot 5^4 \cdot 7}C_2^3C_8 - \tfrac{1429571}{2^{13} \cdot 3^8 \cdot 5^9 \cdot 7}C_2^7$ \\[5pt]
        \bottomrule
    \end{tabular}
    \end{threeparttable}
    \caption{The degree $(p+1)$ potentials obtained as Chevalley lifts of polynomial twisted periods for various $p$ and $\mathfrak g$ satisfying $\gcd(p,h^\vee)=1$. When $p < h^\vee$, these are precisely the lifts of the Saito flat basic invariants.}
    \label{tbl:potentials}
\end{table}

The exceptional entries in Table \ref{tbl:potentials} are useful because the construction takes place first on a Cartan subalgebra of rank six, seven, or eight. Once the invariant polynomial has been found there, Chevalley restriction determines its lift to the full Lie algebra. We therefore do not need a separate coefficient search in 78, 133, or 248 adjoint coordinates to determine the potential. The Jacobian ideal is nevertheless taken in those adjoint coordinates, or in the coordinates of the chosen Slodowy slice after Higgsing. The reduction to the Cartan simplifies the construction of $P$; it does not replace the protected ring by a ring of Weyl-invariant functions on the Cartan.
This gives a finite algebraic prescription for every admissible $p$, even when the resulting expression is too long to display conveniently. Explicit exceptional flat coordinates are available in the literature, including the $E_7$ and $E_8$ constructions \cite{Abriani,Talamini:2010jg}.

\subsection{The Higgs branch and \texorpdfstring{$P$}{P}} \label{sec:HB}

We now show that the degree $(p+1)$ potentials
constructed from the polynomial twisted periods satisfy the Higgs branch condition in equations \eqref{eqn:BIGHIGGS} and \eqref{eqn:HiggsAD}. That is:
\begin{equation}
    \operatorname{Crit}(P) = \overline{\mathcal{O}}_p \,.
\end{equation}
Looking at the definitions, we thus need to show that
\begin{equation}\label{eqn:HBHB}
    \nabla P(X) = 0 \quad \Longleftrightarrow \quad (\operatorname{ad}_X)^{2p} = 0 \,,
\end{equation}
for all $X \in \mathfrak{g}_\mathbb{C}$. If $(\operatorname{ad}_X)^{2p} = 0$ then $X$ is nilpotent. For $X$ nonzero, we can choose an $\mathfrak{sl}_2$-triple $(X, H, Y)$, with $[H, X] = 2X$. Since $P$ is a $G_\mathbb{C}$-invariant degree $(p+1)$ polynomial, then $\nabla P$ is equivariant and has degree $p$, and thus
\begin{equation}
    [H, \nabla P(X)] = 2 p \nabla P(X) \,.
\end{equation}
Decomposing $\mathfrak g_{\mathbb{C}}$ into irreducible modules for this $\mathfrak{sl}_2$, the condition $(\operatorname{ad}_X)^{2p}=0$ implies that every summand has highest $H$-weight strictly less than $2p$. Consequently, there is no nonzero vector of $H$-weight $2p$, and hence $\nabla P(X)=0$. At $X=0$, the same conclusion follows because $\nabla P$ is homogeneous of positive degree $p$. Thus
\begin{equation}\label{eqn:bianca}
     \overline{\mathcal{O}}_p \subseteq \operatorname{Crit}(P) \,.
\end{equation}

To show the opposite implication in equation \eqref{eqn:HBHB}, we first show that if $P$ is the Chevalley lift of a polynomial twisted period, then $\nabla P(X) = 0$ implies that $X$ is nilpotent. It is enough to exclude nonzero semisimple critical points. Indeed, if $X=X_s+X_n$ is the Jordan decomposition, the semisimple part $X_s$ belongs to the closure of the adjoint orbit of $X$. Since the critical locus of an invariant polynomial is closed and invariant under conjugation, $\nabla P(X)=0$ implies $\nabla P(X_s)=0$. If $X_s \neq 0$, conjugating $X_s$ into the Cartan gives some nonzero $x \in \mathfrak{h}$, which then gives a critical point of $q=P|_{\mathfrak{h}}$.

Let us define $F = \nabla q$, and let us suppose that  $x\in\mathfrak{h}$ is a non-zero critical point of $q$, thus $F(x) = 0$. Since $F$ is a non-zero polynomial map on the Cartan, the Taylor expansion around $x$ must have a first non-zero term, say at order $m$:\footnote{The integer $m$ is the order of vanishing of $F$ at $x$; it is unrelated to the earlier notation $c/2$.}
\begin{equation}\label{eq:blue-Taylor}
	F(x+ty)=t^m f_m(y)+O(t^{m+1}),\qquad m\geq1,\qquad f_m\not\equiv0 \,.
\end{equation}
From the chain rule, we have that
\begin{equation}
    \tfrac{d}{dt} F(x + ty) = H_q(x + ty)y \,,
\end{equation}
and since the polynomial twisted period relation, in equation \eqref{eqn:ptpdef}, holds for every $y$, we can rewrite this as a sum over the roots. We obtain
\begin{equation}\label{eqn:cake}
    m t^{m-1} f_m(y) + \mathcal{O}(t^m) = \frac{p}{h^\vee} \sum_{\alpha \in \Phi^+} \frac{ \langle \alpha, y \rangle \langle \alpha, F(x + ty)\rangle}{\langle \alpha, x \rangle + t \langle \alpha, y \rangle} \alpha \,.
\end{equation}
Since the numerator on the RHS is $\mathcal{O}(t^m)$, the coefficient of $t^{m-1}$ on the LHS can only receive contributions from roots $\alpha$ such that $\langle \alpha, x \rangle = 0$. Define
\begin{equation}\label{eq:blue-Psi}
	\Psi_x=\{\alpha\in\Phi: \langle \alpha,x \rangle=0\},\qquad \Psi_x^+=\Psi_x\cap\Phi^+ \,.
\end{equation}
These are the roots of the semisimple part of the centralizer of $x$. They form a proper parabolic root subsystem.\footnote{For complex $x$, the parabolic assertion can be justified without assuming $x$ lies in a real Weyl chamber. Write $x=a+ib$ in the real form of the Cartan. A generic real linear combination of $a$ and $b$ has exactly the same vanishing roots as $x$. Conjugate that real vector to a closed Weyl chamber; its vanishing roots form a standard parabolic subsystem. The root spans have a nondegenerate restriction of the bilinear form because they are complexifications of Euclidean root spaces.} The roots in $\Psi_x^+$ are thus precisely those whose denominator term $ \langle \alpha,x \rangle $ in the twisted period equation vanishes at $x$.  In equation \eqref{eqn:cake}, we keep $y$ generic, with $ \langle \alpha,y \rangle \neq0$ for $\alpha\in\Psi_x$ and $f_m(y) \neq 0$.

Extracting the $t^{m-1}$ coefficients from the left and the right of equation \eqref{eqn:cake}, we obtain the following identity:
\begin{equation}\label{eq:blueoperator}
	m f_m(y)=\frac{p}{h^\vee}A_{\Psi_x}(f_m(y)),\qquad A_{\Psi_x}(v)=\sum_{\alpha\in\Psi_x^+}\langle \alpha,v \rangle \alpha \,.
\end{equation}
Since this identity is polynomial in $y$ it holds for all $y$. It is the analogue, for the roots that vanish at $x$, of the root-sum identity used in equation \eqref{eq:twisted}. Now decompose the parabolic root subsystem into irreducible components,
$\Psi_x=\Psi_{x,1}\sqcup\cdots\sqcup\Psi_{x,s}$, and let $h_a$ be the Coxeter number of $\Psi_{x,a}$. Since $x\neq0$, every component is proper and thus satisfies $0<h_a<h^\vee$. The root-sum operator $A_{\Psi_x}$
acts by $h_a$ on the span of $\Psi_{x,a}$ and by zero on the orthogonal complement of the span of $\Psi_x$. The component of $f_m$ in the orthogonal complement must vanish because $m>0$. Any nonzero component in the span of $\Psi_{x,a}$ would require
\begin{equation}
	m h^\vee=p h_a \,.
\end{equation}
Coprimality of $p$ and $h^\vee$ then implies $h^\vee\mid h_a$, which is impossible for a proper component. Consequently we would require $f_m=0$, contradicting its definition. Only for $x = 0$ does the properness assumption fail, allowing $f_m \neq 0$.

Thus $q$ has no critical point other than the origin. The preceding Jordan decomposition argument then shows that every critical point of $P$ is nilpotent. Combining this with equation \eqref{eqn:bianca}, we find
\begin{equation}\label{eq:blue-conclusion}
    \overline{\mathcal{O}}_p\subseteq\Crit(P)\subseteq\mathcal{N} \,.
\end{equation}
For $p>h^\vee$, the orbit closure $\overline{\mathcal{O}}_p$ is the full nilpotent cone, so the two inclusions prove the desired equality. For $p<h^\vee$, they establish nilpotency of the support; equality with $\overline{\mathcal{O}}_p$ additionally requires excluding the nilpotent orbits outside this closure.

We now take $X \in \operatorname{Crit}(P)$. For both $A$ and $D$ type $\mathfrak{g}_\mathbb{C}$, the universal potential $P$ is given by the determinant formula (up to normalization)
\begin{equation}
   P(X) = [t^{p+1}] \, \operatorname{det}(\bm{1} - t X)^{p/h^\vee} \,,
\end{equation}
where we recall that the determinant, and thus the trace below, are taken in the defining representation of $\mathfrak{g}_\mathbb{C}$. Therefore, for any $Y \in \mathfrak{g}_\mathbb{C}$, we have 
\begin{equation}
    (\nabla P(X), Y) = \frac{d}{ds}(P(X + sY))|_{s=0} = - \frac{p}{h^\vee}\,[t^{p+1}]\,\operatorname{det}(\bm{1} - tX)^{p/h^\vee}\operatorname{tr}(t(\bm{1} - tX)^{-1}Y) \,.
\end{equation}
Since $X$ is a point on the critical locus of $P$, we can use the just-proven fact that $X$ must be nilpotent. This implies that
\begin{equation}
    \operatorname{det}(\bm{1} - tX) = 1 \,, \qquad (\bm{1} - tX)^{-1} = \sum_{k \geq 0} t^k X^k \,,
\end{equation}
and therefore
\begin{equation}
    0 = \langle \nabla P(X), Y \rangle = - \frac{p}{h^\vee} \operatorname{tr}(X^p Y) \,.
\end{equation}
We can see that $X^p \in \mathfrak{g}_\mathbb{C}$ for both $A$ (since $X$ is nilpotent thus $X^p$ is traceless) and $D$ (since $p$ is odd, by $\gcd(p, h^\vee) = 1$, the skew-symmetry is preserved) type algebras, and thus non-degeneracy of the trace pairing implies that $X^p = 0$. Using the definition of the adjoint action in terms of commutators, we can then write
\begin{equation}
    \operatorname{ad}_X^{2p}(Y) = \sum_{k=0}^{2p} (-1)^k \binom{2p}{k} X^{2p - k} Y X^k \,.
\end{equation}
Since $X^p = 0$, then $\operatorname{ad}_X^{2p}$ vanishes as well. This argument concludes that $\operatorname{Crit}(P) \subseteq \overline{\mathcal{O}}_p$, and thus, in combination with equation \eqref{eqn:bianca}, we have proven that 
\begin{equation}
    \operatorname{Crit}(P) = \overline{\mathcal{O}}_p \,,
\end{equation}
whenever $\mathfrak{g}_\mathbb{C}$ is $\mathfrak{sl}_N(\mathbb{C})$  or $\mathfrak{so}_{2N}(\mathbb{C})$. For the theories with $\mathfrak{g}$ exceptional and $p < h^\vee$, there are exactly fifteen theories that satisfy the necessary condition of $\gcd(p, h^\vee) = 1$. It is straightforward to determine the polynomial twisted periods $P$ explicitly, and thus to check that $\nabla P$ is nonzero on representatives of every nilpotent orbit outside $\overline{\mathcal{O}}_p$. We have verified this explicitly in all fifteen cases, establishing the remaining inclusion in each case, and hence $\operatorname{Crit}(P)=\overline{\mathcal{O}}_p$ for the exceptional theories.

\section{Properties of the universal potential} \label{sec:Punique}

Thus far, we have constructed a potential $P$ for $\mathcal{D}_p(G)$, given by the Chevalley lift of a polynomial twisted period, which satisfies the Higgs branch condition in equation \eqref{eqn:BIGHIGGS} and admits a syzygy in bidegree $(p+c-1,1)$ of the first jet scheme. We now study some of the properties of the quotient ring $R_P = \mathbb{C}[\mathfrak{g}_\mathbb{C}] \, / \, J_P$.

\subsection{Minimality}\label{sec:minimality}

In this section, we argue that this syzygy is \emph{minimal}, that is, it contributes to the algebraic Betti numbers as in equation \eqref{eqn:BIGSYZYGY}. Specifically, we want to show that the singlet syzygy
\begin{equation}\label{eqn:targetSYZYGY}
    \mathcal{I} = (c-1) \langle \nabla C_c(X), F_1(X) \rangle - p \langle H_{C_c}(X)Y, F_0(X) \rangle = 0 \,,
\end{equation}
where $C_c$ is the Chevalley lift of the degree $c$ Saito flat basic invariant $t_c$, $F_0(X) = \nabla P(X)$, and $F_1(X) = H_P(X)Y$\footnote{Here we relabeled $X_0 \to X$, $X_1 \to Y$ in Section \ref{sec:unip} to avoid clutter.}, cannot be derived from the existence of a syzygy in a lower bidegree. Throughout, we regard a syzygy as its coefficient vector with respect to the defining relations $(F_0,F_1)$; equalities between syzygies are understood in this module. Since the homogeneous maximal ideal is generated by $X$ and $Y$, if $\mathcal{I}$ is non-minimal then it can be written as
\begin{equation}
    \mathcal{I} = \langle X, \mathcal{T} \rangle + \langle Y, \mathcal{U} \rangle \,,
\end{equation}
where $\mathcal{T}$ and $\mathcal{U}$ are, not necessarily minimal, syzygies of bidegrees:
\begin{equation}
    \operatorname{deg}(\mathcal{T}) = (p+c-2,1) \,, \qquad \operatorname{deg}(\mathcal{U}) = (p+c-2,0) \,.
\end{equation}
Since $\mathcal{I}$ is $G_\mathbb{C}$-invariant, then $\mathcal{T}$ and $\mathcal{U}$ can be chosen to be adjoint-valued and $G_\mathbb{C}$-equivariant.

Now we can consider the usual strategy, that we have used throughout this section, of restricting to the Cartan:
\begin{equation}
  \begin{gathered}
    X|_\mathfrak{h} = x \,, \quad Y|_\mathfrak{h} = y \,, \quad P(X)|_\mathfrak{h} = q(x) \,, \\ F_0(X)|_\mathfrak{h} = f_0(x) = \nabla q(x) \,, \quad F_1(X)|_\mathfrak{h} = f_1(x) = H_q(x) y \,.
  \end{gathered}
\end{equation}
And then the restriction of $\mathcal{I}$ to the Cartan becomes
\begin{equation}
    \mathcal{I}_\mathfrak{h} = \langle x, \mathcal{T}_\mathfrak{h} \rangle +  \langle y, \mathcal{U}_\mathfrak{h} \rangle \,,
\end{equation}
in the obvious notation. We can see that $\mathcal{U}_\mathfrak{h}$ is a syzygy of jet degree zero; however, since the Higgs branch condition in equation \eqref{eqn:BIGHIGGS} implies that $\operatorname{Crit}(q) = \{0\}$, the syzygy module is generated by the Koszul syzygies. For all theories satisfying $\gcd(p, h^\vee) = 1$, we have $p > c-2$, and therefore
\begin{equation}
    \operatorname{deg}(\mathcal{U}_\mathfrak{h}) = p + c - 2 < 2p \,,
\end{equation}
which is incompatible with $\mathcal{U}_\mathfrak{h}$ being Koszul. Therefore, $\mathcal{U}_\mathfrak{h}$ vanishes, and we are left with
\begin{equation}\label{eqn:gnastygnorc}
    \mathcal{I}_\mathfrak{h} = \langle x, \mathcal{T}_\mathfrak{h} \rangle \,.
\end{equation}
We now prove that there is no $\mathcal{T}_\mathfrak{h}$ that can reproduce the specific form of our proposed minimal syzygy $\mathcal{I}_\mathfrak{h}$.

Since $\mathcal{I}_\mathfrak{h}$ is Weyl-invariant, we can take $\mathcal{T}_\mathfrak{h}$ to be a Weyl-equivariant family of syzygies. Let $\{h_i\}$ be an orthonormal basis of the Cartan, then we have a collection of syzygies
\begin{equation}\label{eqn:putativelowersyzygies}
    (\mathcal{T}_\mathfrak{h})_i = \langle a_i(x,y), f_0 \rangle + \langle b_i(x), f_1 \rangle = 0 \,,
\end{equation}
for each $i = 1, \cdots, \operatorname{rank}(\mathfrak{g})$. The bidegrees of the polynomials $a_i$ and $b_i$ are constrained by the required bidegree of $\mathcal{T}_\mathfrak{h}$: $\operatorname{deg}(a_i) = (c-2,1)$ and $\operatorname{deg}(b_i) = (c-2,0)$. In this basis, we can write $x = \sum_i x_i h_i$, and then
\begin{equation}
    \langle x, \mathcal{T}_\mathfrak{h} \rangle = \left\langle\sum_i x_i a_i(x,y), f_0\right\rangle + \left\langle\sum_i x_i b_i, f_1\right\rangle = 0 \,.
\end{equation}
If this were to reproduce the syzygy $\mathcal{I}_\mathfrak{h}$, then we would need that
\begin{equation}\label{eqn:abreq}
    \sum_i x_i a_i(x, y) = -pH_{t_c}(x)y \,, \qquad \sum_i x_i b_i(x) = (c-1) \nabla t_c(x) \,.
\end{equation}

We first consider the case of $\mathfrak{g} = \mathfrak{su}(N \geq 3)$, in which case we have $c = 3$. The degrees and Weyl-equivariance of $a_i$ and $b_i$ require that they take the form
\begin{equation}
    a_i(x,y) = \eta \sum_{j,k} d_{i,j,k}y_k h_j \,, \qquad b_i(x) = \lambda \sum_{j,k} d_{i,j,k}x_k h_j \,,
\end{equation}
where $d_{i,j,k}$ spans the one-dimensional space of Weyl-invariant three-tensors on $\mathfrak{h}$. If we normalize so that
\begin{equation}
    t_c(x) = \frac{1}{6} \sum_{i,j,k} d_{i,j,k} x_i x_j x_k \,,
\end{equation}
then matching with equation \eqref{eqn:abreq} forces $\eta = -p$ and $\lambda = 1$. We now show that the lower degree syzygies as in equation \eqref{eqn:putativelowersyzygies} involving these $a_i$ and $b_i$ violate the Higgs branch assumption in equation \eqref{eqn:BIGHIGGS}. Consider the following symmetric matrices:
\begin{equation}
    H_{i,j} = \partial_i \partial_j q \,, \quad K_{i,j} = \sum_k d_{i,j,k}x_k \,, \quad M_{i,j} = \sum_k d_{i,j,k}\partial_k q \,.
\end{equation}
In terms of these matrices, the putative syzygy in equation \eqref{eqn:putativelowersyzygies} becomes
\begin{equation}
    \langle a_i, \nabla q \rangle +  \langle b_i, H_q y\rangle  = \sum_j \left( \sum_\ell K_{i,\ell} H_{\ell,j} - p M_{i,j} \right)y_j \,.
\end{equation}
Assuming that this syzygy exists, the RHS must vanish for all $y_j$ and we conclude that
\begin{equation}
    KH = pM \,.
\end{equation}
Since the matrices are symmetric, we can take the transpose to see that $HK = KH$, and thus we observe that $[K,M] = 0$. Plugging in the explicit expressions, we find that 
\begin{equation}
    [K,M]_{i,j} = \sum_{a,k,\ell} \left( d_{i,a,k}d_{a,j,\ell} - d_{i,a,\ell}d_{a,j,k} \right) x_k \partial_\ell q \,.
\end{equation}
Since the cubic invariant tensor satisfies the following contraction identity:
\begin{equation}
    \sum_{a} \left( d_{i,a,k}d_{a,j,\ell} - d_{i,a,\ell}d_{a,j,k} \right) \,\,=\,\, \kappa (\delta_{i,\ell}\delta_{j,k} - \delta_{i,k}\delta_{j,\ell} )\,,
\end{equation}
for some non-zero normalization $\kappa$. Consequently, the commutator is
\begin{equation}
    [K, M]_{ij} \,\,=\,\, \kappa( x_j \partial_i q - x_i \partial_j q ) \,.
\end{equation}
For $i \neq j$, these are syzygies of degree $(p+1, 0)$, which violates 
the implication of the Higgs branch condition
that $\mathbb{C}[\mathfrak h]/J_q$ is a complete
intersection, and thus all jet-degree-zero syzygies are Koszul (and thus of degree at least $(2p,0)$). Hence, a $\mathcal{T}_\mathfrak{h}$ such that $\mathcal{I}_\mathfrak{h} = \langle x, \mathcal{T}_\mathfrak{h} \rangle $ does not exist, and thus $\mathcal{I}_\mathfrak{h}$ is minimal. Thus the syzygy in equation \eqref{eqn:targetSYZYGY} is minimal.

We can make a similar argument for $\mathfrak{g} = \mathfrak{so}(2N)$, with $N \geq 5$. The degree of the next-to-minimal fundamental invariant is $c = 4$. From the degrees of $a_i$ and $b_i$, we see that the former is bilinear in $x$ and $y$, and the latter is quadratic in $x$. The Weyl-invariant four-tensors are spanned by the symmetric tensor $d_{i,j,k,\ell}$, as well as $\delta_{i,j}\delta_{k,\ell}$, $\delta_{i,k}\delta_{j,\ell}$, and $\delta_{i,\ell}\delta_{j,k}$. Therefore, we have the generic ansatz:
\begin{equation}
    \begin{aligned}
        a_i(x,y)
        &= \eta \sum_{j,k,\ell} d_{i,j,k,\ell} x_k y_\ell h_j
        + \gamma \sum_j x_j y_i h_j
        + \delta \sum_j x_i y_j h_j
        + \varepsilon \sum_{j,k} \delta_{i,j} x_k y_k h_j \,, \\
        b_i(x)
        &= \lambda \sum_{j,k,\ell} d_{i,j,k,\ell} x_k x_\ell h_j
        + \alpha \sum_{j,k} \delta_{i,j} x_k^2 h_j
        + \beta \sum_j x_i x_j h_j \,.
    \end{aligned}
\end{equation}
The $\eta$, $\gamma$, $\delta$, $\varepsilon$, $\lambda$, $\alpha$, and $\beta$ are seven undetermined constants. We define the symmetric four-tensor associated to the prescribed Saito-flat invariant by
\begin{equation}
    t_c = \frac{1}{24} \sum_{i,j,k,\ell} d_{i,j,k,\ell} x_i x_j x_k x_\ell \,,
\end{equation}
and then use the matching condition in equation \eqref{eqn:abreq} to solve for most of these coefficients. In the end, we find
\begin{equation}
    \begin{aligned}
        a_i(x,y)
        &= -\frac{p}{2}\sum_{j,k,\ell} d_{i,j,k,\ell} x_k y_\ell h_j
        + \gamma \sum_j
        \left( x_j y_i - \delta_{i,j}\sum_k x_k y_k \right) h_j \,, \\
        b_i(x)
        &= \frac{1}{2}\sum_{j,k,\ell} d_{i,j,k,\ell} x_k x_\ell h_j
        + \alpha \sum_j
        \left( \delta_{ij}\sum_k x_k^2 - x_i x_j \right) h_j \,.
    \end{aligned}
\end{equation}
Unlike the $\mathfrak{su}(N)$ case, there are free coefficients left over after imposing the matching condition. As before, we define the matrices
\begin{equation}
    H_{i,j} = \partial_i\partial_j q \,, \qquad
    K_{i,j} = \frac{1}{2}\sum_{k,\ell} d_{i,j,k,\ell}x_kx_\ell \,, \qquad
    M_{i,j} = \sum_{k,\ell} d_{i,j,k,\ell}x_k\partial_\ell q \,.
\end{equation}
Using Euler homogeneity,
\begin{equation}
    \sum_\ell x_\ell H_{\ell, j} = p\,\partial_j q \,,
\end{equation}
the putative lower-degree syzygies in equation \eqref{eqn:putativelowersyzygies} become
\begin{equation}\label{eqn:soputsyz}
    \begin{aligned}
        0 &= \langle a_i,\nabla q \rangle + \langle b_i,H_qy \rangle \\
        &= \sum_j \Biggl[
        \sum_\ell K_{i\ell}H_{\ell j}
        -\frac{p}{2}M_{ij}
        +\alpha\left(\sum_k x_k^2\right)H_{ij}
        -p\alpha x_i\partial_jq \\
        &\hspace{70mm}
        +\gamma\delta_{ij}\sum_k x_k\partial_kq
        -\gamma x_j\partial_iq
        \Biggr]y_j \,.
    \end{aligned}
\end{equation}
Since this expression must vanish for every $y$, each coefficient of $y_j$ must vanish. Multiplying these coefficient equations by $x_j$ and summing over $j$, we can use
\begin{equation}
    \sum_j H_{i,j}x_j = p\,\partial_i q \,, \qquad
    \sum_j M_{i,j}x_j
    = 2\sum_\ell K_{i,\ell}\partial_\ell q \,,
\end{equation}
to obtain
\begin{equation}
    (p\alpha-\gamma)
    \left[
        \left(\sum_k x_k^2\right)\partial_i q
        -x_i\sum_j x_j\partial_j q
    \right]=0 \,.
\end{equation}
If $p\alpha-\gamma$ were nonzero, this would give a nonzero coefficient vector defining a syzygy among the zeroth jet relations $\partial_j q$, of bidegree $(p+2,0)$. Since $p\geq3$, we have $p+2<2p$, and such a syzygy is excluded by the complete intersection property of the zeroth jet scheme. Consequently,
\begin{equation}
    \gamma=p\alpha \,.
\end{equation}
We have one free coefficient, $\alpha$, and we now show that there is no constant $\alpha$ such that equation \eqref{eqn:soputsyz} vanishes for all $i$, as would be necessary for the decomposition of our target syzygy as in equation \eqref{eqn:gnastygnorc}. We define the symmetric matrix
\begin{equation}
    R_{i,j}
    = \frac{1}{2}M_{i,j}
    + \alpha\left(
        x_i\partial_j q+x_j\partial_i q
        -\delta_{i,j}\sum_k x_k\partial_k q
    \right) \,.
\end{equation}
The coefficient equations can then be written as
\begin{equation}
    \left(K+\alpha\left(\sum_k x_k^2\right)\mathbf{1}\right)H
    = pR \,.
\end{equation}
Since $H$, $K$, and $R$ are symmetric, transposing gives
\begin{equation}
    H\left(K+\alpha\left(\sum_k x_k^2\right)\mathbf{1}\right)
    = pR \,.
\end{equation}
Comparing these equations yields $[K,H]=0$. Consequently,
\begin{equation}
    p[K,R]
    = \left[K,
        \left(K+\alpha\left(\sum_k x_k^2\right)\mathbf{1}\right)H
      \right]
    = 0 \,.
\end{equation}
Then, we can write
\begin{equation}
    0 = [K, R]_{i,j} = \sum_{k} C_{i,j}^k(x; \alpha) \partial_k q  \,,
\end{equation}
where the $C_{i,j}^k$ are cubic polynomials in $x$. Any non-zero such polynomial would lead to a syzygy of the zeroth jet scheme, but since (for $p > 3$)
\begin{equation}
    \operatorname{deg}(C_{i,j}^k \partial_k q) = p + 3 < 2p \,,
\end{equation}
then this syzygy is not Koszul, and thus it would be incompatible with the Higgs branch condition. Let us consider this case first: we start by considering the coefficients $C_{i,j}^j$ with $i \neq j$. We choose the normalizations:
\begin{equation}
    t_4 = \sum_{k} x_k^4 - \frac{3}{2(N-1)}s^2 \,, \qquad s = \sum_k x_k^2 \,,
\end{equation}
and thus we can write
\begin{equation}\label{eqn:gulp}
    K_{i,j} = A x_i^2 \delta_{i,j} + \frac{B}{2}s\delta_{i,j} + B x_i x_j \,, \quad \qquad A = 12 \,, \quad B = - \frac{12}{N-1} \,.
\end{equation}
We can then extract the coefficient of $\partial_j q$ in $[K,R]_{i,j}$, and we find that it is
\begin{equation}\label{eqn:Cijj}
    C_{i,j}^j = x_i \left[AB x_j^2 + \left(\frac{B}{2} + \alpha \right)(A(x_i^2 - x_j^2) + Bs) \right] \,,
\end{equation}
where we emphasize that the $j$ is not summed over. Choose $\ell \neq i, j$. Then equation \eqref{eqn:Cijj} contains a monomial 
\begin{equation}
    B\left(\frac{B}{2} + \alpha \right)x_i x_\ell^2 \,.
\end{equation}
For $C_{i,j}^j$ to vanish, the coefficient of this monomial must be zero, and since $B$ is non-zero, we have $\alpha = -B/2$. Substituting this into equation \eqref{eqn:Cijj}, we obtain
\begin{equation}
    C_{i,j}^j = ABx_i x_j^2 \,,
\end{equation}
which is non-zero. Therefore, there is no choice of the coefficient $\alpha$ which causes $C_{i,j}^j$ to vanish as a polynomial, and thus $\mathcal{I}_\mathfrak{h}$ is a minimal syzygy for $p > 3$. Finally, for the case where $p = 3$, we can use the fact that $q = t_4$ directly in the putative lower syzygy in equation \eqref{eqn:soputsyz}, and we see explicitly that the coefficient of $y_j$, for $i \neq j$, is
\begin{equation}
    x_i x_j \left[
        A\left(\frac{B}{2}-\alpha\right)(x_i^2+x_j^2)
        +\left(\frac{B^2}{2}-2B\alpha\right)s
    \right]\,,
\end{equation}
for the same $A$, $B$ as in equation \eqref{eqn:gulp}. 
Again, by looking at the individual monomials, there exists no $\alpha$ for which this is the zero polynomial. Thus, for $\mathfrak{g} = \mathfrak{so}(2N)$ and $N \geq 5$, we have shown that the syzygy $\mathcal{I}$ in equation \eqref{eqn:targetSYZYGY} is minimal.

We have checked the remaining four cases, $\mathfrak{g} = \mathfrak{so}(8), \mathfrak{e}_6, \mathfrak{e}_7, \mathfrak{e}_8$. While the arguments have the same form, the analysis gets somewhat lengthy since the $a_i$ and $b_i$ have many more terms as $c$ increases. We refrain from writing these cases explicitly here.

\subsection{Uniqueness}\label{sec:unique}

We are considering degree-$(p+1)$ $G_\mathbb{C}$-invariant polynomials $P$ which solve equation \eqref{eqn:BIGPDE}, with $C_c$ taken to be the Chevalley lift of the degree-$c$ Saito polynomial, $t_c$.\footnote{For even $c$, this choice of $C_c$ in equation \eqref{eqn:BIGPDE} is an additional restriction beyond the existence of the syzygy, as discussed in Section \ref{sec:nonunique}.} Section \ref{sec:ptp} established that the Chevalley lift of a polynomial twisted period with twisting parameter $p/h^\vee$ provides such a solution. We now prove that it is the unique solution, up to overall normalization.

Let $q = P|_\mathfrak{h}$ be a homogeneous Weyl-invariant polynomial of degree $p+1$, and let $t_c$ be the prescribed Saito polynomial of degree $c$. Suppose that $P$ solves equation \eqref{eqn:BIGPDE}. Restriction to the Cartan then gives
\begin{equation}\label{eq:fixed}
	(c-1)H_q(x)\nabla t_c-pH_{t_c}(x)\nabla q=0 \,.
\end{equation}
Write $K=H_{t_c}$ for convenience. Differentiating equation \eqref{eq:fixed} and antisymmetrizing the two free indices cancels the third derivatives and gives
\begin{equation}\label{eq:commuting}
	(p+c-1)[H_q,K]=0.
\end{equation}
We can thus conclude that $H_q$ commutes with $K$. Next, we define the operator, $H_0(x)$, that the Hessian $H_q(x)$ would be equal to if $q$ were a polynomial twisted period:
\begin{equation}
	H_0(x) u=\frac{p}{h^\vee}\mathcal{B}_x \langle u,\nabla q \rangle \,,
\end{equation}
where $\mathcal{B}_x(u,v)$ is as defined in equation \eqref{eqn:defCalB}. We now prove that $H_q(x) = H_0(x)$.
Since $t_c$ is a polynomial twisted period, we have
\begin{equation}
    K(x)u=\frac{c-1}{h^\vee}\mathcal{B}_x \langle u,\nabla t_c \rangle \,.
\end{equation}
The almost-dual Coxeter multiplication $\mathcal{B}_x$ is commutative and associative on the regular Cartan; its multiplication operators commute \cite{MR2999308}. 
Therefore, we can conclude that 
\begin{equation}
 [H_0,K]=0 \,.
\end{equation}
We also know that the root-sum identity and Euler homogeneity give
\begin{equation}
	H_0(x)x=p\nabla q(x)=H_q(x)x \,.
\end{equation}
Then, we have determined that $D(x)=H_q(x)-H_0(x)$ commutes with $K(x)$ and annihilates $x$. It follows that 
\begin{equation}
	D(x)(K(x)^j x) \,\,=\,\,K(x)^j(D(x)x)\,\,=\,\,0 \qquad \text{for all } j\geq 0 \,.
\end{equation}
Thus, if 
\begin{equation}\label{eqn:stilton}
    x,\,K(x)x,\,\cdots,\,K(x)^{r-1}x \,,
\end{equation}
span the $r$-dimensional Cartan subalgebra on a nonempty Zariski-open set, then $D=0$ there, and since the entries of $D$ are rational functions the vanishing extends throughout the regular Cartan. 
Feigin--Silantyev \cite{MR2999308} prove that,
for $\gcd(p,h^\vee)=1$, the nonconstant homogeneous
polynomial twisted period of degree $p+1$ and
twisting parameter $p/h^\vee$ is unique up to
normalization. Consequently, the solution $P$
of equation \eqref{eqn:BIGPDE}, with $C_c$ fixed
to be the lift of the Saito polynomial, is also
unique up to normalization.

\subsection{Non-uniqueness}\label{sec:nonunique}

In Section \ref{sec:syzygyeqn}, we argued that the existence of $G_\mathbb{C}$-invariant polynomials, $P$ and $C_c$ of degrees $(p+1)$ and $c$ respectively, satisfying equation \eqref{eqn:BIGPDE}, would imply the existence of a bidegree $(p-1+c,1)$ syzygy in the first jet ring: $J_1(R_P)$. In Section \ref{sec:ptp}, we showed that if we take $C_c$ to be the Chevalley lift of the Saito polynomial $t_c$, and $P$ the Chevalley lift of the polynomial twisted period of degree $(p+1)$ and twisting parameter $p/h^\vee$, then the pair $(P, C_c)$ solves equation \eqref{eqn:BIGPDE}. We further showed that the resulting syzygy is minimal in Section \ref{sec:minimality}, and thus it contributes to the algebraic Betti numbers as in equation \eqref{eqn:BIGSYZYGY}.

When the degree, $c$, of the next-to-minimal fundamental invariant of $\mathfrak{g}_\mathbb{C}$ is \emph{odd}, then $C_c(X)$ is a unique $G_\mathbb{C}$-invariant polynomial, up to an overall normalization. In particular, $C_c$ is the Chevalley lift of the Saito polynomial of degree $c$, $t_c$. Therefore, we can use the uniqueness arguments of Section \ref{sec:unique} to show that equation \eqref{eqn:BIGPDE} forces $P$ to be the lift of the polynomial twisted period with twisting parameter $p/h^\vee$. 

When $c$ is \emph{even}, the situation is rather more complicated. Let $C_2(X)$ and $C_c(X)$ belong to a chosen set of homogeneous generators of $\mathbb{C}[\mathfrak g_{\mathbb C}]^{G_{\mathbb C}}$. Then, modulo overall normalization, there exists a family of degree $c$ $G_\mathbb{C}$-invariant polynomials:
\begin{equation}
    C_c + a C_2^{c/2} \,.
\end{equation}
For generic $a$, these invariant polynomials are \emph{not} Chevalley lifts of the Saito polynomial $t_c$, and thus they do not arise from a polynomial twisted period. The generic syzygy condition is in equation \eqref{eqn:cevengensyzygy}; without loss of generality, we can take $C_c$ to be the Chevalley lift of the degree $c$ Saito polynomial, $t_c$, by redefining the parameter $\mu$. For $\mu = 0$, the same uniqueness argument for $c$ odd then forces $P$ to be a lift of the polynomial twisted period with twisting parameter $p/h^\vee$. However, we can also consider $P$ that solve equation \eqref{eqn:cevengensyzygy} with $\mu \neq 0$. Such $P$ are generally not lifts of polynomial twisted periods, and lead to non-isomorphic quotient rings, even though they can also satisfy the Betti number condition in equation \eqref{eqn:BIGSYZYGY} and the Higgs branch condition in equation \eqref{eqn:BIGHIGGS}.

To illustrate this, let us consider the example of $\mathcal{D}_3(SO(10))$. The generic form of the syzygy condition is equation \eqref{eqn:cevengensyzygy}. This condition has multiple solutions. For example, let
\begin{equation}
    P_1(X) = \frac{1}{2} \operatorname{tr}_{\bm{10}} X^4 - \frac{3}{32}(\operatorname{tr}_{\bm{10}} X^2)^2 \,, \qquad P_2(X) = \frac{1}{2} \operatorname{tr}_{\bm{10}} X^4 \,.
\end{equation}
The potential $P_1$ provides a syzygy of the form in equation \eqref{eqn:cevengensyzygy} with $\mu = 0$, whereas $P_2$ provides a syzygy with $\mu = -9$. Note that only $P_1$ is the Chevalley lift of a polynomial twisted period. While
\begin{equation}
    \operatorname{Crit}(P_1) = \operatorname{Crit}(P_2) = \overline{\mathcal{O}}_{[3^3,1]} \,,
\end{equation}
which is the correct Higgs branch for this SCFT, the quotient rings by the associated Jacobian ideals are different, as graded $G_\mathbb{C}$-algebras:
\begin{equation}
    R_1 = \mathbb{C}[\mathfrak{so}(10)_\mathbb{C}]  \, / \, J_{P_1} \quad\not\cong\quad R_2 = \mathbb{C}[\mathfrak{so}(10)_\mathbb{C}] \, / \, J_{P_2} \,.
\end{equation}
Nevertheless, it appears that the Hilbert series of the arc spaces of these rings are identical
\begin{equation}
    \operatorname{HS}_{q,q}(J_\infty(R_1)) \quad = \quad \operatorname{HS}_{q,q}(J_\infty(R_2))\,,
\end{equation}
at least to the order which we have been able to verify. In this sense, the Schur index and the Higgs branch do not provide sufficient information to select which of these non-isomorphic rings should be associated to $\mathcal{D}_3(SO(10))$. To shine a light on this problem, we turn to the associated VOA, which is $\mathbb{V} = L_{-16/3}(\mathfrak{so}(10))$, and consider Zhu's $C_2$-algebra $R_\mathbb{V} = \mathbb{V} \, / \, C_2(\mathbb{V})$. Doing the explicit singular vector calculation, we find that there is a cubic adjoint relation in $R_\mathbb{V}$, which is
\begin{equation}
    X^3 - \frac{3}{16}(\operatorname{tr}_{\bm{10}} X^2)X = 0 \,.
\end{equation}
This is proportional to $\nabla P_1$, and therefore the VOA selects the Jacobian ideal associated to the potential given by the uplift of a polynomial twisted period.

We can also consider the example of $\mathcal{D}_5(SO(10))$, for which we construct a family of degree $6$ potentials
\begin{equation}
    P_b = \frac{1}{2} \operatorname{tr}_{\bm{10}} X^6 + \frac{5}{8} b (\operatorname{tr}_{\bm{10}} X^2)(\operatorname{tr}_{\bm{10}} X^4) + \frac{25}{144} b^2 (\operatorname{tr}_{\bm{10}} X^2)^3  \,,
\end{equation}
that satisfy the syzygy condition in equation \eqref{eqn:cevengensyzygy} with parameter
\begin{equation}
    \mu = -40b - 15 \,.
\end{equation}
We can compute the critical locus for generic $b$, and we find that
\begin{equation}
    \operatorname{Crit}(P_b) = \overline{\mathcal{O}}_{[5^2]} \quad \text{ for } \quad b \not\in \left\{ -\frac{6}{5} ,  -\frac{3}{5},  -\frac{2}{5},  -\frac{3}{10} ,  -\frac{6}{25},  -\frac{1}{5},  -\frac{3}{20},  -\frac{3}{25}\right\} \,.
\end{equation}
That is, all except a small finite set of values of $b$ reproduce the correct Higgs branch. As for $\mathcal{D}_3(SO(10))$, we can pass to the VOA, compute the singular vector, and project onto the $C_2$-algebra, and we observe the following degree five adjoint relation:
\begin{equation}
    X^5 - \frac{5}{16} (\operatorname{tr}_{\bm{10}} X^2) X^3 - \frac{5}{32} (\operatorname{tr}_{\bm{10}} X^4) X + \frac{25}{512} (\operatorname{tr}_{\bm{10}} X^2)^2 X = 0 \,.
\end{equation}
Up to overall normalization, this is 
\begin{equation}
    \nabla P_{-3/8} \,,
\end{equation}
which corresponds to a syzygy with $\mu = 0$. Again, the additional information of the vertex operator algebra fixes the potential precisely to the Chevalley lift of a polynomial twisted period. Based on these examples, and a general notion of commonality between the odd and even $c$ theories, we would propose that the syzygy maximization in the even $c$ case should be augmented by the condition that $\mu$ in equation \eqref{eqn:cevengensyzygy} should vanish. Then $P$ is necessarily the lift of a polynomial twisted period, for all $\mathcal{D}_p(G)$ theories. 

\section{The Macdonald index}\label{sec:Macdonald}

In \cite{Kang:2026nge,Kang:2026ona}, it was proposed that the Macdonald index for certain 4d $\mathcal{N}=2$ SCFTs could be reproduced from the Hilbert series of the arc space of certain filtered graded affine schemes. For Argyres--Douglas theories, we proposed that the coordinate rings of these affine schemes were obtained as quotients of a polynomial ring by a Jacobian ideal of a universal potential, $P$. Furthermore, we argued that this quotient must satisfy a syzygy-maximization condition, as given here in equation \eqref{eqn:BIGSYZYGY}. In Section \ref{sec:unip}, we argued that if $P$ is the Chevalley lift of a polynomial twisted period, then it satisfies both the syzygy-maximization condition and the Higgs branch condition. All that remains is to define the $T$-filtration, and hence we can now compute the Hilbert series of the arc space, as in equation \eqref{eq:indexproposal}, to obtain predictions for the Macdonald indices. These indices are heretofore unknown, but we can verify that the Schur and Hall--Littlewood limits \cite{Gadde:2011uv} are reproduced.

\subsection{The \texorpdfstring{$T$}{T}-filtration}

The proposal in equation \eqref{eq:indexproposal} is that the Macdonald index is reproduced by computing the Hilbert series of the arc space of a filtered graded affine scheme $\operatorname{Spec}(R_P)$:
\begin{equation}
    \operatorname{HS}_{q,q,T}(\operatorname{gr} J_\infty(R_P)) \,.
\end{equation}
Thus far, we have only discussed the construction of the $q$-graded ring
\begin{equation}
    R_P = \mathbb{C}[\mathfrak{g}_\mathbb{C}] \, / \, J_P \,,
\end{equation}
where the potential $P$ is given by the Chevalley lift of the degree-$(p+1)$ polynomial twisted period with twisting parameter $p / h^\vee$. We can equip $R_P$ with a compatible $T$-filtration as follows. Let $\mathfrak{n}$ denote the nilradical of $R_P$, that is
\begin{equation}
    \mathfrak{n} = \{ f \in R_P \mid f^N = 0 \text{ for some } N \geq 1 \} \,.
\end{equation}
We can then define a filtration on each graded piece of $R_P$ as follows. Let $(R_P)_d$ denote the homogeneous component of $R_P$ of degree $d$, and define
\begin{equation}\label{eqn:RFILT}
    F_s(R_P)_d = \bigl(\mathfrak{n}^{\,\max\{d-s,0\}}\bigr)_d \,, \qquad s \in \mathbb{Z} \,,
\end{equation}
with $\mathfrak{n}^0 = R_P$.  This is an increasing filtration:
\begin{equation}
    F_s(R_P)_d \subseteq F_{s+1}(R_P)_d \,,
    \qquad
    F_s(R_P)_d \, F_r(R_P)_e
    \subseteq F_{s+r}(R_P)_{d+e} \,.
\end{equation}
We propose that this $T$-filtration is precisely the $T$-filtration such that
\begin{equation}\label{eqn:ARCy}
    I_\text{Mac}(q,T) = \operatorname{HS}_{q,q,T}(\operatorname{gr} J_\infty(R_P)) \,,
\end{equation}
where the coordinate ring of the arc space carries the induced multiplicative filtration, with jet derivatives preserving filtration degree. Here, and henceforth, $\operatorname{gr}$ is the associated graded ring with respect to this nilradical filtration.

The Hall--Littlewood index \cite{Gadde:2011uv} is obtained from the Macdonald index by taking the limit $q \rightarrow 0$ with $t = qT$ fixed. This picks out terms of the form $q^n T^n$ in the Macdonald index. For $\mathcal{D}_p(G)$ theories it is expected that the Hall--Littlewood index is identical to the Hilbert series of the Higgs branch \cite{Gadde:2011uv,Kang:2022zsl}.
The nilradical filtration in equation \eqref{eqn:RFILT} automatically ensures that the Hall--Littlewood limit of the proposed arc space Hilbert series reproduces the Higgs branch Hilbert series. Indeed, let $(d,m,s)$ denote polynomial degree, jet degree, and filtration degree, respectively. After taking the jet scheme, the filtration satisfies $s\leq d$, while $m\geq0$. Upon setting $T=t/q$, a homogeneous contribution becomes $q^{d+m}T^s=q^{d+m-s}t^s$, which survives as $q\to0$ only if $m=0$ and $s=d$. Thus only the diagonal of the jet-degree-zero algebra contributes; all positive jets are suppressed. Since the jet-degree-zero algebra is $R_P$, its surviving component is
\begin{equation}\label{eqn:HLlimit}
  (\operatorname{gr} R_P)_{d,d} = \frac{F_d(R_P)_d}{F_{d-1}(R_P)_d} = \frac{(R_P)_d}{\mathfrak n_d} = (R_P/\mathfrak n)_d \,.
\end{equation}
Using $R_P/\mathfrak n\cong\mathbb{C}[\overline{\mathcal{O}}_p]$, we therefore obtain
\begin{equation}
  \lim_{\substack{q\to0\\qT=t}}
  \operatorname{HS}_{q,q,T}\left(\operatorname{gr} J_\infty(R_P)\right)
  =\operatorname{HS}_t\left(\mathbb{C}[\overline{\mathcal{O}}_p]\right) \,.
\end{equation}
This identification holds at the level of graded algebras and hence includes all relations and syzygies. For theories whose Hall--Littlewood index equals their Higgs branch Hilbert series, the proposal therefore reproduces this limit by construction. Note that recovering the Higgs branch coordinate ring on the diagonal requires lowering the filtration degree of every nilpotent element; the nilradical filtration in equation \eqref{eqn:RFILT} is the minimal multiplicative filtration implementing this requirement.\footnote{Minimal in the following sense. We could consider an alternative filtration where we take some non-generating element of the nilradical and assign it a lower $T$-degree than that induced by its expression in terms of those generators. Both the Schur and Hall--Littlewood limits are, in general, blind to such a change. In the nilradical filtration we have defined in equation \eqref{eqn:RFILT}, we avoid adding any such extra conditions.}

Let us now compare the nilradical filtration to the filtration structure we discussed in \cite{Kang:2026ona}. Consider the case where $\mathfrak{g}_\mathbb{C} = \mathfrak{sl}_N(\mathbb{C})$. For any value of $p \geq 2$, recalling also that $\gcd(p,N) = 1$, the nilradical is the ideal generated by a choice of primitive invariants
\begin{equation}
    \mathfrak{n} = (C_2, \cdots, C_N) \subset R_P \,.
\end{equation}
Note that we are using the same notation for a polynomial in $\mathbb{C}[\mathfrak{g}_\mathbb{C}]$ and its image in $R_P$. Since $C_i(X) \in \mathfrak{n}$, we have that $C_i(X) \in F_{i-1}(R_P)_i$. Therefore, we can define auxiliary coordinates $y_2, \cdots, y_N$ with bidegree $\operatorname{deg}(y_i) = (i, i-1)$, and write
\begin{equation}
    R_P \cong \frac{\mathbb{C}[\mathfrak{g}_\mathbb{C}] \otimes_\mathbb{C} \mathbb{C}[y_2, \cdots, y_N]}{(J_P, y_2 - C_2(X), \cdots, y_N - C_N(X))} \,,
\end{equation}
The relation $y_i=C_i(X)$ is homogeneous in the first degree, but inhomogeneous in the second degree. The quotient carries the induced filtration from the grading with respect to the second degree on the underlying polynomial ring. That is, the invariant $C_i(X)$ has a representative of filtration degree $i-1$, even though its homogeneous polynomial expression in the adjoint coordinates has degree $i$. This is the purpose of the auxiliary invariant polynomial factor.

When we have arbitrary ADE $\mathfrak{g}_\mathbb{C}$ with $p > h^\vee$ and $\gcd(p, h^\vee) = 1$, the reduced critical locus is the full nilpotent cone. Since its defining ideal is also generated by the fundamental invariants, we have again
\begin{equation}
  \mathfrak{n}=(C_{d_1},\cdots,C_{d_r}) \subset R_P \,.
\end{equation}
The filtered graded ring therefore admits an auxiliary Casimir presentation similar to that described for $\mathfrak{sl}_N(\mathbb{C})$ above. In fact, we can describe this very generally as follows. Consider first the polynomial ring
\begin{equation}
    \mathbb{C}[\mathfrak{g}_\mathbb{C}] \otimes_\mathbb{C} \mathbb{C}[\mathfrak{g}_\mathbb{C}]^{G_\mathbb{C}} \,,
\end{equation}
and let $x_a$ be the generators of the first factor, and $C_i$ be the generators of the second factor. We assign bidegrees as follows:
\begin{equation}\label{eqn:discbideg}
    \operatorname{deg}(x_a \otimes 1) = (1,1) \,,\qquad \operatorname{deg}(1 \otimes C_i) = (d_i, d_i - 1) \,,
\end{equation}
where $d_i$ is the polynomial degree of the $C_i$. We can then define the multiplication map
\begin{equation}
    \begin{aligned}
        \mu \,:\,\, \mathbb{C}[\mathfrak{g}_\mathbb{C}] \otimes_\mathbb{C} \mathbb{C}[\mathfrak{g}_\mathbb{C}]^{G_\mathbb{C}} &\rightarrow \mathbb{C}[\mathfrak{g}_\mathbb{C}] \,,\\
        f \otimes b &\mapsto fb \,,
    \end{aligned}
\end{equation}
and thus the following filtered graded ring (where the filtration is induced by the second grading in the bidegree assignment in equation \eqref{eqn:discbideg}):
\begin{equation}\label{eq:Rgeneral}
    R = \frac{(\mathbb{C}[\mathfrak{g}_\mathbb{C}] / J_P) \otimes_\mathbb{C} \mathbb{C}[\mathfrak{g}_\mathbb{C}]^{G_\mathbb{C}}}{\pi(\operatorname{ker} \mu)} \,,
\end{equation}
where $\pi$ is the quotient map:
\begin{equation}
    \pi \,:\,\, \mathbb{C}[\mathfrak{g}_\mathbb{C}] \otimes_\mathbb{C} \mathbb{C}[\mathfrak{g}_\mathbb{C}]^{G_\mathbb{C}}  \rightarrow (\mathbb{C}[\mathfrak{g}_\mathbb{C}] / J_P) \otimes_\mathbb{C} \mathbb{C}[\mathfrak{g}_\mathbb{C}]^{G_\mathbb{C}} \,.
\end{equation}

However, when $p < h^\vee$ and $\mathfrak{g}_\mathbb{C} \neq \mathfrak{sl}_N(\mathbb{C})$, the nilradical can have generators which are \emph{not} just the Casimir invariants, and then it is necessary to first determine the nilradical and use the filtered graded structure in equation \eqref{eqn:RFILT} directly. In practice, we can extend the auxiliary variable approach to all theories. Let $(\mathfrak{n}_1, \cdots, \mathfrak{n}_\ell)$ be a homogeneous generating set for the nilradical and let $d_i$ denote the homogeneous polynomial degree of $\mathfrak{n}_i$. Then introduce a factor of a polynomial ring in $y_i$ of $(q,T)$-bidegree $(d_i, d_i - 1)$ together with the relation $y_i - \mathfrak{n}_i$. We will see several examples of this below.

\subsection{Examples}

To observe the effect of the minimal syzygy captured by the Betti numbers in equation \eqref{eqn:BIGSYZYGY}, we should compute the index up to and including order $q^{p+c}$. In this section, we summarize some of the examples where we have computed the Hilbert series of the arc space up to this order, and check that the resulting expressions are consistent with the Macdonald index of the corresponding theory. For $\mathfrak{g}=\mathfrak{su}(N)$, many explicit Macdonald indices appeared in \cite{Kang:2026ona}, and thus we focus on $\mathfrak{g}\neq\mathfrak{su}(N)$ here. An updated list of indices for further examples
will be maintained on the webpage \cite{Atlas}.

As our first example, we can consider the theory $\mathcal{D}_5(SO(10))$. As we can read off from the determinant formula in equation \eqref{eqn:det}, or from Table \ref{tbl:potentials}, the potential is
\begin{equation}
    P(X) = \operatorname{tr}_{\bm{10}} X^6 - \frac{15}{32} (\operatorname{tr}_{\bm{10}} X^2)(\operatorname{tr}_{\bm{10}} X^4) + \frac{25}{512} (\operatorname{tr}_{\bm{10}} X^2)^3 \,,
\end{equation}
where we have written the traces explicitly, and recall that overall rescaling of $P$ does not change the resulting Jacobian ideal. To determine the nilradical $\mathfrak{n}$ of $R_P$, we proceed as follows. The radical of $J_P$ is the ideal of polynomials that vanish on the reduced zero-locus, that is, $\operatorname{Crit}(P)$, which we know is the closure of a nilpotent orbit, in this case $\overline{\mathcal{O}}_{[5,5]}$. Therefore, the nilradical is
\begin{equation}
    \mathfrak{n} = I(\overline{\mathcal{O}}_{[5,5]}) \, / \, J_P \subseteq R_P \,.
\end{equation}
For this nilpotent orbit the ideal has been determined in \cite{MR4126306}, and therefore we can see explicitly that
\begin{equation}
    \mathfrak{n} = (C_2, C_4, \operatorname{Pf}) \,,
\end{equation}
where $\operatorname{Pf}$ is the degree five Pfaffian invariant.
The Hilbert series of the arc space, as in equation \eqref{eqn:ARCy}, of the corresponding filtered graded ring can be determined. Thus, we find the following expression for the expected Macdonald index of $\mathcal{D}_5(SO(10))$:
\begin{equation}\label{eqn:D5SO10Mac}
    \begin{aligned}
        &I_\text{Mac}(q,T) = 1 + 45 T q + (46 T + 1034 T^2)q^2 + (46 T + 2069 T^2 + 16170T^3)q^3 \\ 
    &\quad+ (46 T + 3150 T^{2} + 47520 T^{3} + 193544 T^{4})q^{4} \\ 
    &\quad+ (46 T + 4185 T^{2} + 96119 T^{3} + 742831 T^{4} + 1890578 T^{5})q^{5} \\ 
    &\quad+ (46 T + 5266 T^{2} + 159944 T^{3} + 1858561 T^{4} + 8887842 T^{5} + 15693061 T^{6})q^{6} \\ 
    &\quad+ (46 T + 6301 T^{2} + 240973 T^{3} + 3703967 T^{4} \\&\qquad\qquad\qquad\qquad\,\,\, + 26322117 T^{5} + 86787190 T^{6} + 113806506 T^{7})q^{7} \\ 
    &\quad+ (46 T + 7382 T^{2} + 337228 T^{3} + 6488807 T^{4} + 60849117 T^{5} \\&\qquad\qquad\qquad\qquad\,\,\,+ 295309838 T^{6} + 720138022 T^{7} + 735756825 T^{8})q^{8} \\
    &\quad+ (46 T + 8417 T^{2} + 450733 T^{3} + 10375187 T^{4} + 121240078 T^{5} + 773537660 T^{6} \\&\qquad\qquad\qquad\qquad\,\,\,+ 2755599588 T^{7} + 5220589739 T^{8} + 4305782382 T^{9})q^{9} + \cdots \,.
    \end{aligned}
\end{equation}
No expression for the Macdonald index of this theory was previously known. We can test the validity of this expression by matching to the known limits of the Macdonald index. In particular, the $T = 1$ limit reproduces the Schur index as in equation \eqref{eqn:Schurgeneral}. As we have said, the Higgs branch of $\mathcal{D}_5(SO(10))$ is the closure of the $\mathfrak{so}(10)$ nilpotent orbit associated to the D-partition $[5,5]$. The Hilbert series for this symplectic singularity was worked out in \cite{Cabrera:2017ucb} where it was found to be
\begin{equation}
   \begin{aligned}
    \operatorname{HS}(t) &= 1 + 45t + 1034t^2 + 16170t^3 + 193544t^4
+ 1890578t^5 \\
&\quad\,\,\,\,\,+ 15693061t^6 + 113806506t^7 + 735756825t^8
+ 4305782382t^9  + \cdots \,.
  \end{aligned}
\end{equation}
It is clear to see that this matches the Hall--Littlewood limit of equation \eqref{eqn:D5SO10Mac}, as expected from equation \eqref{eqn:HLlimit}.

Next, we consider the example of $\mathcal{D}_5(SO(8))$; $\mathfrak{g} = \mathfrak{so}(8)$ a priori appeared exceptional since the Betti number condition in equation \eqref{eqn:BIGSYZYGY} showed a jump of two minimal syzygies instead of one. Regardless, the polynomial twisted period analysis goes through, and we obtain the potential written in Table \ref{tbl:potentials}. As in the $\mathcal{D}_5(SO(10))$, the nilradical can be worked out from the vanishing ideal of the critical locus, $I(\overline{\mathcal{O}}_{[5,3]})$, which was given in \cite{MR4126306}. The result is again an ideal generated by the primitive invariants:
\begin{equation}
    \mathfrak{n} = (C_2, C_4, \operatorname{Pf}) \,.
\end{equation}
Since $p + c = 9$, we compute the Hilbert series of the arc space to at least this order, and we find the following conjectural Macdonald index:
\begin{equation}
    \begin{aligned}
    &I_\text{Mac}(q,T) = 1 + 28 T q + (29 T + 405 T^{2})q^2 + (29 T + 811 T^{2} + 4032 T^{3})q^3 \\ 
    &\quad+ (29 T + 1246 T^{2} + 11719 T^{3} + 31057 T^{4})q^{4} \\ 
    &\quad+ (29 T + 1652 T^{2} + 23870 T^{3} + 116577 T^{4} + 197232 T^{5})q^{5} \\ 
    &\quad+ (29 T + 2087 T^{2} + 39704 T^{3} + 291998 T^{4} + 897347 T^{5} + 1074537 T^{6})q^{6} \\ 
    &\quad+ (29 T + 2493 T^{2} + 59975 T^{3} + 581100 T^{4} \\&\qquad\qquad\,\, + 2641086 T^{5} + 5695114 T^{6} + 5159675 T^{7})q^{7} \\ 
    &\quad+ (29 T + 2928 T^{2} + 83929 T^{3} + 1019000 T^{4} + 6077090 T^{5} \\&\qquad\qquad\,\,+ 19108982 T^{6} + 31006408 T^{7} + 22266063 T^{8})q^8 \\ 
    &\quad+ (29 T + 3334 T^{2} + 112349 T^{3} + 1629016 T^{4} + 12089977 T^{5} + 49593699 T^{6} \\&\qquad\qquad\,\,+116087794 T^{7} + 148767775 T^{8} + 87632983 T^{9})q^9 \\
    &\quad + (29 T + 3769 T^{2} + 144423 T^{3} + 2447512 T^{4} + 21660525 T^{5} + 109899844 T^{6} \\&\qquad\,\,+ 333615077 T^{7} + 611684331 T^{8} + 641380083 T^{9} + 318185643 T^{10})q^{10} + 
    \cdots \,.
  \end{aligned}
\end{equation}
Setting $T=1$ matches the Schur limit obtained from the expansion of equation \eqref{eqn:Schurgeneral}, and the Hall--Littlewood limit reproduces the Higgs branch Hilbert series of $\overline{\mathcal{O}}_{[5,3]}$, given in \cite{Cabrera:2017ucb}, which is indeed the Higgs branch of $\mathcal{D}_5(SO(8))$.

As our first example where there are additional contributions to the nilradical, we can consider $\mathcal{D}_3(SO(10))$. As we can read off from the determinant formula in equation \eqref{eqn:det} the potential is
\begin{equation}
    P(X) = \operatorname{tr}_{\bm{10}} X^4 - \frac{3}{16} (\operatorname{tr}_{\bm{10}} X^2)^2 \,,
\end{equation}
where we have written the traces explicitly. The Higgs branch is the closure of the nilpotent orbit associated to the $[3,3,3,1]$ partition. 
A complete generating set for
$I(\overline{\mathcal O}_{[3,3,3,1]})$ is not
known to us, but explicit computation identifies
the following generators of the nilradical:
\begin{equation}\label{eqn:D3SO10nil}
    \mathfrak{n} = (C_2, \partial_a \operatorname{Pf}, \cdots ) \,,
\end{equation}
by explicit computation. Here the $\cdots$ may include terms of polynomial degree $\geq 7$. For the purposes of our calculation, we consider the filtered graded structure defined by equation \eqref{eqn:RFILT} with $\mathfrak{n}$ replaced by $\mathfrak{n}' = (C_2, \partial_a \operatorname{Pf}) \subseteq \mathfrak{n}$ and then use the known Higgs branch Hilbert series to verify, since $R_P / \mathfrak{n}' \rightarrow R_P / \mathfrak{n}$ is a graded surjection, that
\begin{equation}
    (\mathfrak{n}')_d =  (\mathfrak{n})_d \,,
\end{equation}
for all $d \leq 8$. Thus, using this filtered graded structure and computing the Hilbert series of the arc space as in equation \eqref{eqn:ARCy}, we find a putative expression for the Macdonald index of $\mathcal{D}_3(SO(10))$:
\begin{equation}
  \begin{aligned}
    &I_\text{Mac}(q,T) = 1 + 45 T q + (46 T + 1034 T^2)q^2 + (46 T + 2069 T^2 + 16125T^3)q^3 \\ 
    &\quad+ (46 T + 3150 T^{2} + 47519 T^{3} + 191520 T^{4})q^{4} \\ 
    &\quad+ (46 T + 4185 T^{2} + 96118 T^{3} + 740408 T^{4} + 1844447 T^{5})q^{5} \\ 
    &\quad+ (46 T + 5266 T^{2} + 159943 T^{3} + 1856091 T^{4} + 8780509 T^{5} + 14983010 T^{6})q^{6} \\ 
    &\quad+ (46 T + 6301 T^{2} + 240972 T^{3} + 3701451 T^{4} \\&\qquad\qquad\qquad\qquad\,\,\, + 26148911 T^{5} + 84365161 T^{6} + 105503683 T^{7})q^{7} \\ 
    &\quad+ (46 T + 7382 T^{2} + 337227 T^{3} + 6486245 T^{4} + 60611927 T^{5} \\&\qquad\qquad\qquad\qquad\,\,\,+ 290016579 T^{6} + 683180645 T^{7} + 657069459 T^{8})q^{8} + \cdots \,.
  \end{aligned}
\end{equation}
We note that no such expression has been written down before. By taking the $T \rightarrow 1$ limit, we recover the Schur index for which the closed-form expression was given in equation \eqref{eqn:Schurgeneral}. On the other hand, we can compute the Hall--Littlewood limit and we find:
\begin{equation}\label{eqn:D3SO10HL}
    \begin{aligned}
    I_\text{HL}(t) &= 1 + 45 t + 1034 t^2 + 16125t^3 + 191520 t^{4} + 1844447t^{5} \\ 
    &\qquad\qquad+ 14983010 t^{6} + 105503683 t^{7} + 657069459 t^{8} + \cdots \,.
  \end{aligned}
\end{equation}
As discussed already, the Higgs branch is the closure of the nilpotent orbit associated to the $[3,3,3,1]$ D-partition of $10$. The Hilbert series of this space has been explicitly computed in \cite{Cabrera:2017ucb}, and we find that it is indeed identical to what we find from the arc space calculation in equation \eqref{eqn:D3SO10HL}. This verifies that there is no other generator of the nilradical $\mathfrak{n}$, beyond $C_2$ and $\partial_a \operatorname{Pf}$, that contributes at polynomial degree less than $9$.

Finally, we consider an E-type example, specifically $\mathcal{D}_7(E_6)$. Since $p < h^\vee$, the potential $P$ is simply the Chevalley lift of the degree $8$ Saito polynomial, and thus can be written as
\begin{equation}
    P(X) = \operatorname{tr}_{\bm{27}} X^8 - \frac{7}{24} (\operatorname{tr}_{\bm{27}} X^2)(\operatorname{tr}_{\bm{27}} X^6) + \frac{385}{2^{10} \cdot 3^5} (\operatorname{tr}_{\bm{27}} X^2)^4 \,.
\end{equation}
Furthermore, since $p < h^\vee$, the Higgs branch is not the full nilpotent cone; instead equation \eqref{eqn:HiggsAD} reveals it is the closure of the $E_6(a_3)$ nilpotent orbit of $\mathfrak{e}_6$.\footnote{Here we have used the standard Bala--Carter \cite{bala1976classesI,bala1976classesII} labeling for nilpotent orbits of exceptional algebras.} The ideal $I(\overline{\mathcal{O}}_{E_6(a_3)})$ has been worked out in \cite{MR4126306}; it is
\begin{equation}
    I(\overline{\mathcal{O}}_{E_6(a_3)}) = (C_2, C_5, C_6, \partial_a C_8, \partial_a C_9) \,.
\end{equation}
Modulo $(C_2, C_5, C_6)$ the derivative $\partial_a P$ is proportional to $\partial_a C_8$, and therefore the nilradical is simply
\begin{equation}\label{eqn:D7e6n}
    \mathfrak{n} = (C_2, C_5, C_6, \partial_a C_9) \,.
\end{equation}
In particular, this means that in the associated graded the 78 elements of $R_P$ of the form $\partial_a C_9$ should contribute at $(q,T)$-bidegree $(8,7)$. The sensitivity of the Hilbert series of the arc space to the Betti number condition in equation \eqref{eqn:BIGSYZYGY} first appears at order $q^{12}$, and therefore we should consider the $11$th jet scheme. We can compute the Hilbert series of the arc space as in equation \eqref{eqn:ARCy}, and we find
\begin{align}\label{eqn:D7e6Mac}
        &I_\text{Mac}(q,T) = 1 + 78 T q + (79 T + 3080 T^{2})q^2 + (79 T + 6161 T^{2} + 82082 T^{3})q^3 \nonumber\\ 
    &\quad+ (79 T + 9321 T^{2} + 243242 T^{3} + 1660659 T^{4})q^{4} \nonumber\\  
    &\quad+ (79 T + 12402 T^{2} + 489643 T^{3} + 6481399 T^{4} + 27203175 T^{5})q^{5} 
    \nonumber\\  
    &\quad+ (79 T + 15562 T^{2} + 815202 T^{3} + 16208037 T^{4} + 131110057 T^{5} + 375783329 T^{6})q^{6} \nonumber\\  
    &\quad+ (79 T + 18643 T^{2} + 1225923 T^{3} + 32343156 T^{4} + 390245936 T^{5} + 2147393245 T^{6} \nonumber\\ &\qquad\quad+ 4502077204 T^{7})q^{7} \nonumber\\  
    &\quad+ (79 T + 21803 T^{2} + 1715802 T^{3} + 56629573 T^{4} + 905436535 T^{5} \nonumber\\ &\qquad\quad+ 7388597368 T^{6} + 29659761820 T^{7} + 47746972976 T^{8})q^{8} \nonumber\\  
    &\quad+ (79 T + 24884 T^{2} + 2290922 T^{3} + 90566631 T^{4} + 1806532655 T^{5} + \nonumber\\ &\qquad\quad 19492119110 T^{6} + 115490845922 T^{7} + 355289973289 T^{8} + 455325093300 T^{9})q^{9} \\ 
    &\quad+ (79 T + 28044 T^{2} + 2945121 T^{3} + 135906469 T^{4} + 3245442051 T^{5} \nonumber\\ &\qquad\quad + 43582277436 T^{6} + 340601254442 T^{7} + 1540622717341 T^{8} \nonumber\\ &\qquad\quad + 3767535902988 T^{9} + 3952591273708 T^{10})q^{10} \nonumber\\  
    &\quad+ (79 T + 31125 T^{2} + 3684561 T^{3} + 194139266 T^{4} + 5403577876 T^{5} \nonumber\\ &\qquad\quad+ 86770726209 T^{6} + 842103924262 T^{7}  + 5012007171099 T^{8} \nonumber\\ &\qquad\quad+ 17964543845291 T^{9} + 35923309282739 T^{10} + 31545619645659 T^{11})q^{11} \nonumber\\  
    &\quad+ (79 T + 34285 T^{2} + 4503159 T^{3} + 267020163 T^{4} + 8483936158 T^{5} \nonumber\\ &\qquad\quad + 158584900032 T^{6} + 1837792713820 T^{7} + 13554622792627 T^{8} \nonumber\\ &\qquad\quad + 63782045378332 T^{9} + 186450220222233 T^{10} + 311803854043110 T^{11} \nonumber\\ &\qquad\quad + 233370151200876 T^{12})q^{12} 
    + \cdots \,. \nonumber
\end{align}
This is then the proposal for the Macdonald index of $\mathcal{D}_7(E_6)$. We first notice that this is compatible with the Schur index in equation \eqref{eqn:Schurgeneral} when $T = 1$. The closed-form expression for the Hilbert series of this nilpotent orbit closure was determined in \cite{Hanany:2017ooe}, and its expansion to order $t^{12}$ matches the $q^n T^n$ terms in the putative Macdonald index in equation \eqref{eqn:D7e6Mac}.

\section{\texorpdfstring{\boldmath{$\mathcal{D}_p(G, O)$}}{Dp(G,O)} and Slodowy slice restriction}\label{sec:slod}

In \cite{Kang:2026ona}, we argued that once we know the potential $P$ that underlies $\mathcal{D}_p(SU(N))$, then the pullback of that potential can be used to give the filtered graded ring that underlies an arbitrary nilpotent Higgsing of the $\mathfrak{su}(N)$ flavor symmetry. Thus, we here consider the Argyres--Douglas theories $\mathcal{D}_p(G, O)$, with the same constraints on $p$ and $G$ as above, and where $O$ is an arbitrary nilpotent orbit of $\mathfrak{g}_\mathbb{C}$ such that $\mathcal{D}_p(G, O)$ is an interacting SCFT. 

Let $P$ be the same Chevalley uplift of the polynomial twisted period associated to $\mathcal{D}_p(G)$ above and let $(e, h, f)$ be an $\mathfrak{sl}_2$-triple associated to the nilpotent orbit $O$. We have the inclusion of the Slodowy slice, $\mathcal{S}_e$, inside of the algebra:
\begin{equation}
    \iota_e \,:\,\, \mathcal{S}_e \hookrightarrow \mathfrak{g}_\mathbb{C} \,.
\end{equation}
Using this inclusion, we can define the pullback of $P$ to the Slodowy slice and the corresponding quotient ring by the Jacobian ideal:
\begin{equation}
    P_e = \iota^*_e P \,, \qquad\qquad R_{P_e} = \mathbb{C}[\mathcal{S}_e] \, / \, J_{P_e} \,.
\end{equation}

The Slodowy slice, $\mathcal{S}_e = e + \operatorname{ker}(\operatorname{ad} f)$, carries a natural grading. Let $\{T_a\}$ denote a basis of $\operatorname{ker}(\operatorname{ad} f)$, which is homogeneous with respect to the adjoint action of $h$. A general element of the Slodowy slice can be written in terms of affine coordinates, $y_a$, in this basis:
\begin{equation}
    M = e + \sum_a y_a T_a \,.
\end{equation}
The $y_a$ are the analogs of the $x_a$ in equation \eqref{eqn:CgCpoly}. Since the basis is homogeneous under $h$, we can assign the so-called half-Kazhdan degree to each of the affine coordinates:
\begin{equation}
    [h, T_a] = -\kappa_a T_a \quad\Rightarrow\quad \operatorname{deg}(y_a) = 1 + \frac{\kappa_a}{2} \,.
\end{equation}
Since $P_e$ is homogeneous of degree $p+1$
under this grading, the quotient ring $R_{P_e}$
inherits the grading. Now, let $\mathfrak{n}_e$ denote the nilradical of $R_{P_e}$. We can then define a filtration on each homogeneous component in the half-Kazhdan grading. Let $(R_{P_e})_d$ denote the homogeneous component of $R_{P_e}$ of half-Kazhdan degree $d$, and define the filtration
\begin{equation}\label{eqn:RFILTe}
    F_s(R_{P_e})_d = \bigl(\mathfrak{n}_e^{\,\max\{ \lceil d-s \rceil,0\}}\bigr)_d \,, \qquad s \in \frac{1}{2}\mathbb{Z} \,,
\end{equation}
with $\mathfrak{n}_e^0 = R_{P_e}$. The coordinate ring of the arc space carries the induced multiplicative filtration, with jet derivatives preserving both half-Kazhdan degree and filtration degree and increasing jet degree by one. Our proposal is
then that $R_{P_e}$ together with the $T$-filtration as specified in equation \eqref{eqn:RFILTe} is such that
\begin{equation}
    I_{\mathrm{Mac}}^{\mathcal{D}_p(G,O)}(q,T)
    =
    \operatorname{HS}_{q,q,T}
    \bigl(\operatorname{gr}J_\infty(R_{P_e})\bigr) \,.
\end{equation}
Furthermore, 
\begin{equation}
    \operatorname{Spec}(R_{P_e})_{\mathrm{red}}
    \cong
    \mathcal S_e\cap\overline{\mathcal O}_p \qquad \Rightarrow \qquad\mathcal{M}_\mathcal{H}^{\mathcal{D}_p(G, O)} = \operatorname{Spec}(R_{P_e})_\text{red} \,,
\end{equation}
that is, the Higgs branch of the Higgsed theory follows from the Higgs branch property of the ring $R_P$.

The nilradical $\mathfrak{n}_e$ can be determined directly from $\mathfrak{n}$. Let $\rho:R_P\rightarrow R_{P_e}$ be the surjective ring homomorphism induced by restriction, $\iota_e^*$, to the Slodowy slice. The $G_{\mathbb C}$-invariance of $P$ and Slodowy transversality imply that
\begin{equation}
    \mathfrak{n}_e=\rho(\mathfrak{n}) \,.
\end{equation}
In particular, if $\mathfrak{n}$ is given by the ideal $(f_1, \cdots, f_r)$, where the $f_i$ are polynomial representatives of the nilradical generators, then 
\begin{equation}
    \mathfrak{n}_e = (\iota_e^* f_1, \cdots, \iota_e^* f_r) \,.
\end{equation}
That is, we simply restrict the generators to the Slodowy slice. This reproduces precisely what was found for the nilpotent Higgsing of $\mathcal{D}_p(SU(N))$ in \cite{Kang:2026ona}.

As a fun example, we can consider the $(A_{k-1}, A_{N-1})$ Argyres--Douglas theories, with $\gcd(k,N) = 1$, which are realized as $\mathcal{D}_p(SU(N), [N])$, where $p = k + N$ and $[N]$ denotes the regular nilpotent orbit of $\mathfrak{sl}_N(\mathbb{C})$. In this case, the bigraded polynomial ring becomes:
\begin{equation}\label{eqn:AApoly}
    \mathbb{C}[u_2, \cdots, u_N] \qquad \text{where} \quad \operatorname{deg}(u_a) = (a, a-1) \,.
\end{equation}
Since the reduced critical locus is a point, the nilradical is $\mathfrak{n}_e = (u_2, \cdots, u_N) \subseteq R_{P_e}$, and therefore we can see that the nilradical filtration is induced by assigning bidegree $(a, a-1)$ to $u_a$. The potential, obtained by pulling back $P$ along the Slodowy slice embedding, is simply
\begin{equation}\label{eqn:AApot}
    P_e \propto \operatorname{Res}_{z = \infty} \left( z^N + u_2 z^{N-2} + u_3 z^{N-3} + \cdots + u_N\right)^{\frac{k+N}{N}} dz \,.
\end{equation}
Then the filtered graded ring is simply the quotient of the bigraded polynomial ring in equation \eqref{eqn:AApoly} by the Jacobian ideal associated to the potential $P_e$ in equation \eqref{eqn:AApot}. These potentials exactly match those worked out, for low values of $k$ and $N$, using explicit syzygy-maximization in \cite{Kang:2026nge}. Equation \eqref{eqn:AApot} strongly evokes the Seiberg--Witten geometry of the Coulomb branch of $(A_{k-1}, A_{N-1})$, and a similar expression holds for the $(A, D)$ Argyres--Douglas theories realized as $\mathcal{D}_p(SO(2N), [2N-1,1])$.

\section{Spectral curve and the Hitchin system}\label{sec:Hitchin}

The polynomial appearing in equation \eqref{eqn:AApot} also occurs naturally in the spectral description of the theory. To explain this connection, consider the adjoint quotient
\begin{equation}
	\chi:\mathfrak g_{\mathbb{C}}\longrightarrow\mathfrak h/W\simeq\operatorname{Spec}\mathbb{C}[\mathfrak g_{\mathbb{C}}]^{G_{\mathbb{C}}} \,,
\end{equation}
which assigns to an element of $\mathfrak g_{\mathbb{C}}$ the values of its invariant polynomials. When $e$ belongs to the principal nilpotent orbit, the restriction of $\chi$ to the Slodowy slice $\mathcal{S}_e$ is an isomorphism \cite{Kostant}. Consequently, the restrictions of the basic invariant polynomials provide coordinates on $\mathcal{S}_e$.

For $\mathfrak g_{\mathbb{C}}=\mathfrak{sl}_N(\mathbb{C})$, we choose these coordinates to be the coefficients of the characteristic polynomial
\begin{align}
    F_N(z;u) = \det(z\mathbf 1-X)=z^N+u_2(X)z^{N-2}+\cdots+u_N(X),\qquad X\in \mathcal{S}_e \,.
\end{align}
These are precisely the coordinates $u_2,\ldots,u_N$ used in equation \eqref{eqn:AApoly}, and this characteristic polynomial is the polynomial inside the residue in equation \eqref{eqn:AApot}. Its coefficients are invariant polynomials of degrees $2,\ldots,N$, related to the trace invariants $C_i$ of equation \eqref{eq:CkasTr} by Newton identities.

Hitchin systems provide an integrable-system description of the Coulomb branch geometry of a broad class of 4d $\mathcal{N}=2$ theories \cite{Hitchin1987,Donagi:1995cf}. The same invariant polynomials enter the characteristic equation of a Hitchin field. In that setting, their values vary over the underlying curve and obey the prescribed singularity conditions. This gives a direct connection between the polynomial used to construct our potential and the invariant data entering the spectral curve.

For the principal Higgsing, $\mathcal{D}_p(SU(N), O_\text{prin}) = (A_{k-1}, A_{N-1})$ with $p=N+k$, the protected potential is the residue functional:
\begin{equation}
	P_e(u)\propto\operatorname{Res}_{z=\infty}F_N(z;u)^{(N+k)/N}\,dz \,.
\end{equation}
The fractional power is expanded at infinity as
\begin{equation}
	F_N(z;u)^{(N+k)/N}=z^{N+k}\left(1+\sum_{i=2}^{N}u_i z^{-i}\right)^{(N+k)/N},
\end{equation}
and the residue is minus the coefficient of $z^{-1}$.
Its derivatives are therefore residues of the same polynomial:
\begin{equation}
	\frac{\partial P_e}{\partial u_i}\propto\frac{N+k}{N}\operatorname{Res}_{z=\infty} z^{N-i}F_N(z;u)^{k/N}\,dz \,.
\end{equation}
The familiar curve $w^k+z^N=0$ admits the deformation subfamily 
\begin{equation}
	w^k+F_N(z;u)=0 \,.
\end{equation}
In the $A_{N-1}$ Hitchin description, $w$ is the coordinate on the underlying curve and $z$ is the spectral coordinate. Writing the Higgs field locally as $\Phi=\varphi(w)\,dw$, this subfamily is obtained by choosing
\begin{equation}
	\det\!\left(z\mathbf1-\varphi(w)\right)=z^N+\sum_{i=2}^{N-1}u_i z^{N-i}+u_N+w^k.
\end{equation}
Thus, the characteristic polynomial used in our residue construction appears directly in the Hitchin spectral equation. The natural Seiberg--Witten differential in this presentation is $z\,dw$. The convention $\lambda=w\,dz$ used below is equivalent up to an overall sign and an exact differential, since
\begin{equation}
	w\,dz=-z\,dw+d(zw).
\end{equation}
The full deformation family can also contain mixed monomials in $w$ and $z$ (see \cite{Xie:2012hs}, for example), so this subfamily should not be identified with the entire Coulomb branch family in general.

We determine the Seiberg--Witten scaling dimensions,
denoted by $[\,\cdot\,]_{\mathrm{SW}}$, using
quasi-homogeneity and the normalization
$[\lambda]_{\mathrm{SW}}=1$ for
$\lambda=w\,dz$. The dimensions follow from two requirements. For the curve
\begin{equation}
    w^k+z^N+\sum_{i=2}^N u_i z^{N-i}=0 \,,
\end{equation}
all terms must have the same scaling dimension, so
\begin{equation}
    k[w]_{\mathrm{SW}}=N[z]_{\mathrm{SW}},\qquad [u_i]_{\mathrm{SW}}=i[z]_{\mathrm{SW}} \,.
\end{equation}
The Seiberg--Witten differential $\lambda=w\,dz$ has mass dimension one, giving 
\begin{equation}
    [w]_{\mathrm{SW}}+[z]_{\mathrm{SW}}=1 \,. 
\end{equation}
Solving gives precisely
\begin{equation}
	[z]_{\mathrm{SW}}=\frac{k}{N+k},\qquad [w]_{\mathrm{SW}}=\frac{N}{N+k},\qquad [u_i]_{\mathrm{SW}}=\frac{ik}{N+k} \,.
\end{equation}
These physical dimensions are different from the protected ring assignments $(i,i-1)$. In the spectral family some coefficients are couplings and others are Coulomb-branch expectation values. In the protected construction they are coordinates on the principal slice, subject to the Jacobian relations of $P_e$. It is important to note that the residue potential is a function on this coefficient space; it is not itself the spectral curve equation or the Seiberg--Witten prepotential.

Let us consider an explicit example to concretely demonstrate this distinction. For the $(A_1,A_2)$ SCFT, we have $N=3$ and $k=2$ so that
\begin{equation}
	F_3=z^3+u_2z+u_3,\qquad P_e\propto u_3^2-\frac{1}{9}u_2^3 \,.
\end{equation}
Indeed, the coefficient of $z^{-1}$ in $F_3^{5/3}$ is 
\begin{equation}
    \frac{5}{9}u_3^2-\frac{5}{81}u_2^3 \,.
\end{equation}
The residue at infinity differs from this
coefficient by an overall minus sign, which
does not affect the Jacobian ideal.
Hence we find
\begin{equation}
	\mathbb{C}[u_2,u_3]/J_{P_e}\simeq\mathbb{C}[u_2]/(u_2^2) \,.
\end{equation}
Its reduced spectrum is a point and its nilpotent generator has protected degree $(2,1)$. On the Seiberg--Witten side, the same polynomial gives
\begin{equation}
    [u_2]_{\mathrm{SW}}=\frac{4}{5},\qquad [u_3]_{\mathrm{SW}}=\frac{6}{5} \,,
\end{equation}
i.e., $u_2$ is a coupling and $u_3$ is the Coulomb branch parameter \cite{Xie:2012hs}.

For the exceptional groups, we do not presently have an analogous identification with a deformation subfamily of the Seiberg--Witten geometry. Nevertheless, the common invariant data can be described using the adjoint quotient and the associated cameral cover. Let $C_i$ be homogeneous generators of $\mathbb{C}[\mathfrak g_{\mathbb{C}}]^{G_{\mathbb{C}}}$, with degrees $d_i$. These generators serve two related purposes. In our construction, the potential $P$ is a polynomial in the $C_i$. In a Hitchin system, they describe the gauge-invariant content of a Higgs field varying over a Riemann surface.

To make the latter statement explicit, choose a local coordinate $\xi$ and a local trivialization of the gauge bundle, and write the Higgs field as $\Phi=\varphi(\xi)\,d\xi$. Evaluating the invariant polynomials gives
\begin{equation}
    C_i(\Phi)=b_i(\xi)(d\xi)^{d_i},\qquad b_i(\xi) = C_i(\varphi(\xi))\,,
\end{equation}
where the factor $(d\xi)^{d_i}$ follows from the homogeneity of $C_i$. Hence, the invariant values define meromorphic differentials on the surface. Their allowed poles and leading coefficients are constrained by the puncture data. Once the couplings and masses are fixed, the remaining Coulomb branch deformations are encoded in the allowed variations of these differentials.

Where $\varphi(\xi)$ is regular semisimple, it can be conjugated to an element $x(\xi)\in\mathfrak h$, determined up to the Weyl group.
This Cartan representative is determined only up to the Weyl group $W$. Moreover, continuing it around a closed path can return a different Weyl representative. We can track these choices by passing to a cover of the surface. On a coordinate patch $U$, this cover is called the \emph{cameral cover}, which is defined as \cite{DonagiGaitsgory}
\begin{equation}
    \widetilde{U}=\left\{(\xi,x)\in U\times\mathfrak h\,\middle|\,C_i(x)=b_i(\xi),\quad i=1,\cdots,\operatorname{rank}\mathfrak{g}\right\} \,.
\end{equation}
Its generic fiber consists of the Weyl-related Cartan elements with the prescribed invariant values. For $\mathfrak{sl}_N$, these are the different orderings of the $N$ eigenvalues: the cameral cover generically has $N!$ sheets, while the usual spectral curve has $N$ sheets and records one eigenvalue at a time. The Cartan construction applies equally to the exceptional groups and requires no choice of matrix representation.

In fact, this description makes the common invariant data explicit. For example, in the trace conventions of Table \ref{tbl:potentials}, the $E_6$ potential with $p=5$ is proportional to
\begin{equation}
    \label{eq:PforE6withp5}
    P=C_6-\frac{5}{432}C_2^3 \,.
\end{equation}
Evaluating this polynomial on a Hitchin field gives the degree-six differential
\begin{equation}
    P(\Phi)=\left[ C_6(\varphi(\xi))-\frac{5}{432}C_2(\varphi(\xi))^3\right](d\xi)^6 \,.
\end{equation}
Interestingly, our construction singles out a definite combination of the same invariant differentials that enter the Hitchin description. More precisely, as equation \eqref{eq:PforE6withp5} is the degree-six Saito flat invariant, our construction  selects a particular invariant differential through the twisted period equations.
More generally, the degree $(p+1)$ homogeneity of $P$ gives
\begin{equation}
    P(\Phi)=P(\varphi(\xi))(d\xi)^{p+1} \,.
\end{equation}
This observation follows directly from invariance and homogeneity; identifying a physical role for this particular differential is a further question.

The Coulomb branch effective theory also depends on the Seiberg--Witten differential and its periods. In the type-$A$ spectral description, the differential is locally 
\begin{equation}
    \lambda_{\mathrm{SW}}=y\,d\xi\,,
\end{equation}
where $y$ is an eigenvalue of $\varphi(\xi)$.
The corresponding construction on the cameral cover uses the Cartan-valued one-form $x\,d\xi$, together with the appropriate charge lattice and cycles. Its periods determine the electric and magnetic special coordinates. It would be interesting to determine whether the twisted period defining our exceptional potential can be related to these Seiberg--Witten data. Such a relation would need to explain the specific polynomial selected by the twisted period equations, beyond the appearance of the same invariant generators in both constructions.

\section{Discussion}\label{sec:discussion}

In this paper, we have given a uniform construction of filtered graded affine schemes proposed to encode the Higgs branches and Macdonald indices of the Argyres--Douglas theories $\mathcal{D}_p(G)$, with $G$ of ADE type and $\gcd(p,h^\vee)=1$. The construction relates the syzygy-maximization procedure developed in \cite{Kang:2026nge,Kang:2026ona} to polynomial twisted periods of Coxeter orbit spaces. Their Chevalley lifts define invariant potentials whose Jacobian rings possess the distinguished minimal first-jet syzygies. Fixing the invariant entering the syzygy equation to be the prescribed Saito invariant uniquely selects these potentials, yielding closed determinant expressions for the classical types and a recursive construction for the exceptional potentials.

The reduced critical loci of these potentials reproduce the Higgs branches, while powers of the nilradical supply an intrinsic proposal for the $T$-filtration. The resulting arc space Hilbert series give predictions for Macdonald indices, agreeing with the known Schur and Hall--Littlewood limits in the examples studied. Restriction of the potentials to Slodowy slices extends the construction to nilpotently Higgsed theories, $\mathcal{D}_p(G,O)$. Together, these results provide a uniform prescription for the affine schemes previously sought through theory-by-theory calculations and motivate the following future questions and connections.

\paragraph{\uline{Mathematical conjectures:}} There are a number of standalone mathematical conjectures that follow from our work. One particularly interesting, and perhaps tractable, conjecture is as follows. Let $\mathfrak{g}_{\mathbb{C}}$ be a simple Lie algebra of ADE type, let $p \geq 2$ satisfy $\gcd(p,h^\vee)=1$, and set $k=-h^\vee+h^\vee/p$. Let $(e,h,f)$ be an $\mathfrak{sl}_2$-triple with $e\in\overline{\mathcal O}_p$, and let $\mathcal S_e=e+\ker(\operatorname{ad}f)$ be the corresponding Slodowy slice. We conjecture that Zhu's $C_2$-algebra of the simple $W$-algebra $\mathbb V=\mathcal W_k(\mathfrak{g}_{\mathbb{C}},f)$ is isomorphic, as a graded commutative ring, to
\begin{equation}\label{eqn:RVconj}
    R_{\mathbb V}
    \,\cong\,
    \mathbb{C}[\mathcal S_e]/J_{P_e} \,,
    \qquad
    P_e=P|_{\mathcal S_e} \,,
\end{equation}
where $P$ is the Chevalley lift of the homogeneous polynomial twisted period of degree $p+1$ and twisting parameter $p/h^\vee$, and $J_{P_e}$ is the ideal generated by the derivatives of $P_e$ with respect to affine coordinates on $\mathcal S_e$. The isomorphism identifies the conformal grading on $R_{\mathbb V}$ with the half-Kazhdan grading on the Jacobian ring. The case $e=f=0$ includes the simple affine VOA $L_k(\mathfrak{g}_{\mathbb{C}})$.

\paragraph{\setlength{\ULdepth}{6pt}\uline{Non-simply-laced \boldmath{$\mathcal{D}_p^b(G)$}:}} Polynomial twisted periods also exist for non-simply-laced Coxeter groups, suggesting a potentially straightforward extension of our construction to $\mathcal{D}_p^b(G)$ theories with $G$ of type BCFG. A natural starting point is the subclass whose associated VOAs are simple affine algebras at boundary admissible levels, together with their Drinfeld--Sokolov reductions \cite{Beem:2014rza}. The first question is whether their
$C_2$-algebras admit analogous Jacobian presentations obtained from polynomial twisted periods, thereby generalizing equation \eqref{eqn:RVconj}. Equipping these rings with the nilradical filtration would then provide predictions for their Macdonald indices, which could be tested against independently known
limits. 

\paragraph{\uline{4d mirror symmetry:}} The mirror symmetry proposed in \cite{Shan:2023xtw} relates the associated VOA of a 4d $\mathcal{N}=2$ SCFT to the geometry of its Coulomb branch after compactification on a circle \cite{Fredrickson:2017yka, Fredrickson:2017jcf}. Of particular relevance here is the proposed relation between Zhu's $C_2$-algebra and the cohomology ring of the corresponding Hitchin moduli space, studied there for principal type-$A$ examples. Our explicit Jacobian presentations provide a concrete starting point for testing and extending this relation. Two questions are whether the polynomial twisted periods defining our potentials have a direct interpretation in this correspondence, and whether the proposed $T$-filtration has a geometric counterpart on the Hitchin side.

\paragraph{\uline{Graded unitarity:}}
Graded unitarity provides a formulation of the constraints imposed by 4d unitarity on the associated VOA, involving both an $\mathfrak{R}$-filtration and a compatible conjugation \cite{ArabiArdehali:2025fad,Beem:2026lkq}. It would be interesting to determine whether, for the theories considered here, such a structure induces our proposed nilradical filtration on the $C_2$-algebra, after accounting for the distinction between the $R$-charge and the Macdonald charge $R-r$. A stronger question is whether graded unitarity, together with the prescribed charges of the generators, uniquely selects this filtration, thereby providing a physical characterization of the algebraic prescription proposed here.

\subsection*{Acknowledgements}
M.~J.~K.~and C.~L.~thank the Simons Center for Geometry and Physics for hospitality during the ``23rd Simons Physics Summer Workshop: Theory, Experiment and the Emerging New Physics''. 
We also thank ICTP Trieste and the organizers of the workshop ``Quivers, Symmetries and SCFTs'' for their hospitality when the crucial part of this work was performed. C.~L.~thanks the Yau Mathematical Sciences Center at Tsinghua University for hospitality during the final stages of this work.
We acknowledge the use of generative AI, OpenAI Codex and Anthropic Claude Code for coding. 
M.~J.~K.~is supported by the DOE through DE-SC0010813 and the Start-up Research Grant for new faculty provided by Texas A\&M University.
C.~L.~acknowledges the Deutsche Forschungsgemeinschaft under Germany's Excellence Strategy - EXC 2121 ``Quantum Universe'' - 390833306 and the Collaborative Research Center - SFB 1624 ``Higher Structures, Moduli Spaces, and Integrability'' - 506632645.
The work of J.~S.~is supported by the National Research Foundation of Korea (NRF) grants RS-2024-00405629 and RS-2026-25482546, and the KAIST-KIAS collaboration program. 

\bibliography{references}{}
\bibliographystyle{sortedbutpretty}
\end{document}